\documentclass[11pt]{JHEP3}

\newcommand{\dalpha}{{\dot{\alpha}}}
\newcommand{\dbeta}{{\dot{\beta}}}
\newcommand{\dgamma}{{\dot{\gamma}}}
\newcommand{\ddelta}{{\dot{\delta}}}
\newcommand{\dvarepsilon}{{\dot{\varepsilon}}}
\newcommand{\dkappa}{{\dot{\kappa}}}
\newcommand{\drho}{{\dot{\rho}}}

\newcommand{\CA}{{\cal A}}

\newcommand{\CC}{{\cal C}}

\newcommand{\CE}{{\cal E}}

\newcommand{\CH}{{\cal H}}

\newcommand{\CN}{{\cal N}}
\newcommand{\CO}{{\cal O}}

\newcommand{\CQ}{{\cal Q}}
\newcommand{\CR}{{\cal R}}

\newcommand{\CZ}{{\cal Z}}

\def\13{{\scriptscriptstyle{\!13}}}
\def\7{{\scriptscriptstyle{\!7}}}

\def\R{{\CR}}
\def\RL{{\CH}}
\def\sR{{H}}

\def\R{{\CR}}
\def\Y{{\scriptstyle{Y}}}

\def\B{{B}}

\def\R{{\CR}}

\def\osplus{{\oplus_{\rm{\scriptscriptstyle semi}}}}

\newcommand{\half}{{{\textstyle\frac{1}{2}}}}

\newcommand{\be}{\begin{equation} }
\newcommand{\ee}{\end{equation} }
\newcommand{\ba}{\begin{array}}
\newcommand{\ea}{\end{array}}

\newcommand{\bea}{\begin{eqnarray}}
\newcommand{\eea}{\end{eqnarray}}

\newcommand{\su}{\mbox{su}}
\newcommand{\so}{\mbox{so}}

\def\fB{{B_{\!{\scriptscriptstyle 4}}}}
\def\sB{{B_{\!{\scriptscriptstyle 6}}}}
\def\tB{{B_{\!{\scriptscriptstyle 10}}}}
\def\H{{\stackrel{\rightarrow}{H}}}

\def\tr{{\rm tr}}
\def\Tr{{\rm Tr}}

\newcommand{\stwo}{{{\scriptscriptstyle \!(2,2)}}}
\newcommand{\sfour}{{{\scriptscriptstyle {\rm su}(4)}}}
\newcommand{\ufour}{{{\scriptscriptstyle {\rm u}(4)}}}
\newcommand{\uone}{{{\scriptscriptstyle {\rm u}(1)}}}

\title{Noncentral extension of  the ${\mathbf{AdS_{5}\!\times\!S^{5}}}$ superalgebra~:\\
supermultiplet of  brane charges}

\author{Sangmin Lee${}^{\ast}$ and Jeong-Hyuck Park${}^{\dagger}$\\
\\
${}^{\ast}$CERN, CH-1211, Geneva 23, Switzerland\\
${}^{\dagger}$Institut des Hautes Etudes Scientifiques, Bures-sur-Yvette, 91440, France \\
{}\\
Electronic correspondence: \email{sangmin.lee@cern.ch},~~ \email{park@ihes.fr}}

\abstract{We propose an extension of the $\su(2,2|4)$ superalgebra  
to incorporate  the $F1/D1$ string charges in type IIB string theory on the $AdS_{5}\times S^{5}$ background, or  the 
electro-magnetic charges in the dual super Yang-Mills theory.  
With the  charges introduced,  the superalgebra inevitably  undergoes  a noncentral extension, as noted recently in \cite{Peeters:2003vz}. After developing  a group theoretical method of obtaining the noncentral extension, we show that the  charges form a certain nonunitary  representation of the original unextended superalgebra,  subject to  some constraints. We solve the constraints completely and   show that,  apart from the $\su(2,2|4)$ generators, there exist  $899$ complex brane charges in the extended algebra. Explicitly we   present  all the super-commutation relations among them.  }


\preprint{hep-th/0404051\\
CERN-PH-TH/2004-064\\
IHES/P/04/15\\
STR-04-007}

\begin{document}


\section{Introduction and summary}
$D$-branes have been the cornerstone to understand the non-perturbative aspects of string/M-theory, and the ``central" extensions of super Poincar\'{e} algebras  provide a useful tool to analyze the possible supersymmetric brane configurations.  The  identification of the  central charge with the magnetic charge  of a monopole by Witten and Olive~\cite{Witten:mh}  was the  first crucial step in discovering many exact results in the supersymmetric gauge theories. Also the celebrated  Montonen-Olive duality conjecture~\cite{Montonen:1977sn} received  the first support  from the analysis on the central charges in four dimensional $\CN=4$ super Yang-Mills theory  by Osborn~\cite{Osborn:tq}.  The method has been  applied to the M-theory matrix model on the  flat background~\cite{Banks:1996vh}  by Banks \textit{et al.}~\cite{Banks:1996nn}, and  further to the  pp-wave matrix model~\cite{Berenstein:2002jq} by Hyun and Shin~\cite{Hyun:2002cm} in order to identify all the extended objects. In the supersymmetric field theories  the central charges appear as  surface integrals in the expression of the anti-commutator of the supercharges, while  in the matrix models they come as traces of  a commutator. \newline

Although much effort has been put to obtain the explicit expressions of the brane charges in various theories, it seems that few questions have been addressed to their central property,  which can be, in principle,  straightforwardly checked by investigating the supersymmetry transformations of them.  Historically, the central property was ``proven" in a more abstract way by  Haag,  Lopuszanski and Sohnius~\cite{Haag:1974qh} studying the general structure of  the $Z_{2}$-graded symmetry algebras or the superalgebras. The proof was  based on  the Coleman-Mandula theorem~\cite{Coleman:ad} on all the possible  \textit{symmetry generators} in the  quantum field theories not having trivial scattering amplitudes.   Now  the essential  motivation to question the central property of the brane charges comes from the fact  that   the brane charges   are not symmetry generators nor Noether charges.   Rather, they are  topological living  at the spatial infinity only, and hence free from the constraint by the Haag-Lopuszanski-Sohnius theorem.  In fact, some straightforward manipulations indicate that the generic brane charges are not central.\footnote{Nevertheless all the known solitonic objects seem to have the vanishing values for the novel charges.}\newline

Recently,  Peeters and Zamaklar considered  some extensions of the $AdS$ superalgebra as well as    the  pp-wave superalgebra, and noticed that the brane charges are inevitably non-central~\cite{Peeters:2003vz}  (see  also \cite{Meessen:2003yi} for the related  work). 
The $AdS$ superalgebras are superconformal algebras and  bigger than the super Poincar\'{e} algebras. In particular,  the anti-commutator of the supercharges gives  rotational generators, $M_{ab}$, either for  the anti-de-Sitter space  or for the internal space, under which    the brane charges, say $Z_{a}$,  transform nontrivially. The crucial observation made in \cite{Peeters:2003vz} follows  from the Jacobi identity which contains two supercharges and one brane charge,
\be
{}[\{Q,\,\bar{Q}\},\,Z_{a}]=\{Q,\,[\bar{Q},\,Z_{a}]\}+\{\bar{Q},\,[Q,\,Z_{a}]\}\,.
\ee
By contracting the spinorial indices of the supercharges properly, the left hand side can be set to be an infinitesimal rotation of the brane charge, which do not have any prior reason to vanish. Thus, from the right hand side, one can see  the noncentral property of the brane charge. Namely the brane charge do not commute with the supercharges in general.\newline

In the mathematics literature, all the semi-simple superalgebras were classified by Kac~\cite{Kac:em,Kac:qb} (see also a review by Nahm~\cite{Nahm:1977tg}), but the systematic study of the  noncentral extensions of them remains an open problem. 
The primary goal of the present paper is to explore  the possible  noncentral extensions of the $AdS_{5}\times S^{5}$ superalgebra or $\su(2,2|4)$. There are three types of BPS branes\footnote{For the discussion of the branes on the $AdS$ space, see for example 
\cite{Skenderis:2002vf}.}  one can add  to the anti-commutator of the supercharges, as a starting point for the extension; $F1/D1$ and ${D5/N\!S}5$ charges combine into complex charges, while $D3$ charges are real-valued.  After developing the general method for the extensions, we focus on the electro-magnetic ($F1/D1$) extension. We show that (i) the corresponding extension is unique, (ii) apart from the $\su(2,2|4)$ generators, there are  $899$ complex brane charges in the extended algebra, (iii) the brane charges form a supermultiplet of the original unextended superalgebra, and  we  present  all the super-commutation relations of them explicitly.  Although in the paper we focus on the $AdS_{5}\times S^{5}$ superalgebra,  our method  can be straightforwardly applied to other superalgebras. \newline

The organization as well as the summary of the paper is as follows. 

Section \ref{setting} is to set up our notations to write down the $\su(2,2|4)$ superalgebra in a $\su(2,2)\oplus\su(4)$ covariant way.
In section \ref{u1extension}, we analyze the root structure of the $\su(2,2|4)$ superalgebra and discuss its representations in a self-contained manner. In particular, we focus on a class of representations which are realized by the adjoint actions of  the $\su(2,2|4)$ generators. They  are nonunitary and have finite  dimensions.

Section \ref{main} contains our main results. Motivated by the super Yang-Mills analysis, we define brane charges to be the space integrals of the total derivative terms or the surface integrals. We argue then that the super-commutator involving a brane charge is also a brane charge, and that all the brane charges super-commute with each other. Finally, by investigating  all  possible Jacobi identities, we find out that the brane charges form a ``adjoint representation" of the   original unextended superalgebra,  $\su(2,2|4)$, and that it is subject to some constraints.  In subsection \ref{mainsub}, the constraints are solved completely for the  electro-magnetic extension. We identify the explicit structure of the  supermultiplet and present all the nontrivial super-commutation relations.

In section \ref{comments}, we  describe how to translate our result to the four dimensional language: first for the extended $\CN=4$ superconformal algebra and second for the extended $\CN=4$ super Poincar\'{e} algebra.  We also comment how our extended superalgebra acts on the quantum monopole states  in the super Yang-Mills theory. For the purpose of the last section, in  Appendix  we relate  the  twelve dimensional gamma matrices to the four and ten dimensional ones.

\newpage

\section{$AdS_{5}\times S^{5}$ superalgebra - unextended \label{setting}}
This section is to set up the notations in order to write the  $AdS_{5}\times S^{5}$ superalgebra in terms of the $\su(2,2)\oplus\su(4)$ spinorial conventions. The main formulae are   (\ref{QbarQ}), (\ref{QQ}), (\ref{M1Q}), (\ref{M2Q}), (\ref{MM1}), (\ref{MM2}). 


\subsection{Gamma matrices and spinors}

In order to make the $\mbox{SO}(2,4)\times\mbox{SO}(6)$ isometry of 
$AdS_{5}\times S^{5}$ geometry manifest, it is convenient to employ 
the twelve dimensional gamma matrices of spacetime signature 
$(--++++++++++)$, and write them in terms of two sets of 
six dimensional gamma matrices, $\{\gamma^{\mu}\}$, $\{\gamma^{a}\}$,
\begin{equation}
\begin{array}{ll}
\Gamma^{\mu}=\gamma^{\mu}\otimes\gamma^{(7)}&~~~\mbox{for~}~\mu=1,2,3,4,5,6\\
{}&{}\\
\Gamma^{a}\,=\,1\,\otimes\gamma^{a}&~~~\mbox{for~}~a=7,8,9,10,11,12\,.
\end{array}
\end{equation}
The two sets of the six dimensional gamma matrices satisfy 
\be
\ba{ll}
\gamma^{\mu}\gamma^{\nu}+\gamma^{\nu}\gamma^{\mu}=2\eta^{\mu\nu}\,,~~~&~~~
\gamma^{a}\gamma^{b}+\gamma^{b}\gamma^{a}=2\delta^{ab}\,,
\ea
\ee
where $\eta^{\mu\nu}=\mbox{diag}(--++++)$. With the choice
\begin{equation}
\gamma^{(7)}=
i\gamma^{1}\gamma^{2}\cdots\gamma^{6}=i\gamma^{7}\gamma^{8}\cdots\gamma^{12}
=\left(\begin{array}{cc}1&0\\0&-1\end{array}\right)\,, 
\label{gamma7}
\end{equation}
all the six dimensional gamma matrices  are in the block diagonal form,
\be
\ba{ll}
\gamma^{\mu}=\left(\begin{array}{cc}0&\rho^{\mu}\\ \bar{\rho}^{\mu}&0\end{array}\right)\,,~~~&~~~
\gamma^{a}=\left(\begin{array}{cc}0&\rho^{a}\\ \bar{\rho}^{a}&0\end{array}\right)\,,
\ea
\ee
satisfying the hermiticity conditions,
\be
\ba{ll}
\bar{\rho}_{\mu}=\eta_{\mu\nu}\bar{\rho}^{\nu}=(\rho^{\mu})^{\dagger}\,,~~~&~~~ 
\bar{\rho}^{a}=(\rho^{a})^{\dagger}\,,
\ea\label{herm}
\ee
which ensure that $\Gamma^{1}$, $\Gamma^{2}$ are anti-hermitian and 
others hermitian. 

If we further set all the $4\times 4$ matrices, 
$\rho^{\mu},\,\bar{\rho}^{\mu}$ $\rho^{a},\,\bar{\rho}^{a}$ to
be anti-symmetric~\cite{Park:1998nr}
\begin{equation}
\begin{array}{ll} (\rho^{\mu})_{\alpha\beta}=-(\rho^{\mu})_{\beta\alpha}\,,~~~&~~~
(\bar{\rho}^{\mu})^{\alpha\beta}=
-\half\epsilon^{\alpha\beta\gamma\delta}(\rho^{\mu})_{\gamma\delta}\,,\\
{}&{}\\
({\rho}^{a})_{\dalpha\dbeta}=-({\rho}^{a})_{\dbeta\dalpha}\,,~~~&~~~
(\bar{\rho}^{a})^{\dalpha\dbeta}=
-\half\epsilon^{\dalpha\dbeta\dgamma\ddelta}(\rho^{a})_{\dgamma\ddelta}\,,
\end{array}\label{anti-sym}
\end{equation}
the relations, ${\mbox{su}(2,2)\equiv\mbox{so}(2,4)}$ and 
${\mbox{su}(4)\equiv\mbox{so}(6)}$, become manifest. That is, 
the indices $\alpha,\beta=1,2,3,4$ and $\dalpha,\dbeta=1,2,3,4$  
denote the fundamental representations of 
$\mbox{su}(2,2)$ and $\mbox{su}(4)$, respectively.

It follows that $\{\rho^{\mu}\}$ and  $\{\bar{\rho}^{\mu}\}$ separately form 
bases for the anti-symmetric $4\times 4$ matrices with the 
completeness relation,
\be
\ba{ll}
\tr(\rho^{\mu}\bar{\rho}_{\nu})=4\delta^{\mu}_{~\nu}\,,~~~&~~~
(\rho^{\mu})_{\alpha\beta}(\bar{\rho}_{\mu})^{\gamma\delta}=2(
\delta_{\alpha}{}^{\delta}\delta_{\beta}{}^{\gamma}
-\delta_{\beta}{}^{\delta}\delta_{\alpha}{}^{\gamma})\,.
\ea
\label{antisym}
\ee
On the other hand, the choice of chirality matrices in Eq.(\ref{gamma7}) 
implies that\footnote{We put $\epsilon^{123456}=1$ and ``$[~]$" denotes 
the standard anti-symmetrization with ``strength one".}
\be 
\ba{ll}
\rho^{[\mu}\bar{\rho}^{\nu}\rho^{\lambda]}=
+i\textstyle{\frac{1}{6}}\epsilon^{\mu\nu\lambda\sigma\tau\kappa}
{\rho}_{[\sigma}\bar{\rho}_{\tau}{\rho}_{\kappa]}\,,~~&~~
\bar{\rho}^{[\mu}{\rho}^{\nu}\bar{\rho}^{\lambda]}=
-i\textstyle{\frac{1}{6}}\epsilon^{\mu\nu\lambda\sigma\tau\kappa}
\bar{\rho}_{[\sigma}{\rho}_{\tau}\bar{\rho}_{\kappa]}\,,
\ea
\label{iden}
\ee
so each of the sets
$\rho^{[\mu}\bar{\rho}^{\nu}\rho^{\lambda]}\equiv\rho^{\mu\nu\lambda}$ or 
$\bar{\rho}^{[\mu}{\rho}^{\nu}\bar{\rho}^{\lambda]}\equiv\bar{\rho}^{\mu\nu\lambda}$ 
has only 10 independent components and forms a basis for 
symmetric $4\times 4$ matrices,
\be
\ba{l}
\tr(\rho^{\mu\nu\lambda}
\bar{\rho}_{\sigma\tau\kappa})
=-i4\,\epsilon^{\mu\nu\lambda}{}_{\sigma\tau\kappa}
-24\delta^{[\mu}_{~\sigma}\delta^{\nu}_{~\tau}\delta^{\lambda]}_{~\kappa}\,,\\
{}\\
(\rho^{\mu\nu\lambda})_{\alpha\beta}
(\bar{\rho}_{\mu\nu\lambda})^{\gamma\delta}=-24(
\delta_{\alpha}{}^{\gamma}\delta_{\beta}{}^{\delta}
+\delta_{\beta}{}^{\gamma}\delta_{\alpha}{}^{\delta})\,.

\ea
\label{sym}
\ee
Finally,   $\{\rho^{\mu\nu}\equiv\frac{1}{2}({\rho}^{\mu}\bar{\rho}^{\nu}-{\rho}^{\nu}\bar{\rho}^{\mu})\}$  or $\{\bar{\rho}^{\mu\nu}\equiv\frac{1}{2}(\bar{\rho}^{\mu}{\rho}^{\nu}-\bar{\rho}^{\nu}{\rho}^{\mu})\}$  forms an orthonormal basis for the  general $4\times 4$ traceless matrices,
\begin{equation}
\ba{ll}
\tr(\rho^{\mu\nu}\rho_{\lambda\kappa})=4(\delta^{\mu}{}_{\kappa}\delta^{\nu}{}_{\lambda}
-\delta^{\nu}{}_{\kappa}\delta^{\mu}{}_{\lambda})\,,~&~
\textstyle{-\frac{1}{8}}(\rho^{\mu\nu})_{\alpha}{}^{\beta}(\rho_{\mu\nu})_{\gamma}{}^{\delta}+
\textstyle{\frac{1}{4}}\delta_{\alpha}{}^{\beta}\delta_{\gamma}{}^{\delta}
=\delta_{\alpha}{}^{\delta}\delta_{\gamma}{}^{\beta}\,,
\ea
\label{6dfierz}
\end{equation}
satisfying
\be
(\bar{\rho}^{\mu\nu})^{\alpha}{}_{\beta}=-(\rho^{\mu\nu})_{\beta}{}^{\alpha}\,.\label{rbr}
\ee
Note that precisely the same equations as (\ref{antisym})-(\ref{rbr}) 
hold for the $\mbox{so}(6)$ gamma matrices,  $\{\rho^{a},\,\bar{\rho}^{b}\}$ 
after replacing $\mu,\nu$, $\alpha,\beta$ by 
$a,b$, $\dalpha,\dbeta$, etc. \\

In the above choice of  gamma matrices, 
the twelve dimensional charge conjugation matrices, 
${\CC}_{\pm}$, are  given by 
\begin{equation}
\begin{array}{ll}
\pm(\Gamma^{M}){}^{T}={\CC}_{\pm}\Gamma^{M}{\CC}_{\pm}^{-1}\,,~~M=1,2,\cdots,12,~~&~~~
{\CC}_{\pm}=\left(\begin{array}{cc}0&1\\\pm 1&0\end{array}\right)\otimes\left(\begin{array}{cc}0&1\\\mp 1&0\end{array}\right)\,,
\end{array}
\end{equation}
while the complex conjugate matrices, $\CA_{\pm}$ read
\begin{equation}
\begin{array}{ccc}
\pm(\Gamma^{M}){}^{\dagger}={\CA}_{\pm}\Gamma^{M}{\CA}_{\pm}^{-1}\,,~&~
{{\CA}_{\pm}=\left(\begin{array}{cc}A^{t}&0\\  0&\mp A\end{array}\right)\otimes \left(\begin{array}{cc}1&0\\ 0&\pm 1\end{array}\right)\,,}~&~A=-i\bar{\rho}_{12}=A^{\dagger}=A^{-1}\,.
\ea\label{Apm}
\ee
In particular, for $\mu=1,2,\cdots,6$, we have
\be
\ba{ll}
(\rho^{\mu})^{\dagger}=-A\bar{\rho}^{\mu}A^{t}=\bar{\rho}_{\mu}\,,~~~&~~~
(\bar{\rho}^{\mu})^{\dagger}=-A^{t}\rho^{\mu}A={\rho}_{\mu}\,.
\ea
\ee

Now if we define the twelve dimensional chirality operator as  $\Gamma^{(13)}\equiv\gamma^{(7)}\otimes\gamma^{(7)}$, then
\begin{equation}
\ba{lll}
\{\Gamma^{(13)},\,\Gamma^{M}\}=0\,,~~&~~\CC_{-}=\Gamma^{(13)}\CC_{+}\,,~~&~~ \CA_{-}=\Gamma^{(13)}\CA_{+}\,.
\ea
\ee
In 2+10 dimensions it is possible to impose the Majorana-Weyl condition on spinors  
to have  sixteen independent complex components which coincides with the number of supercharges in the $AdS_{5}\times S^{5}$ superalgebra, ${\mbox{su}(2,2|4)}$.  Up to the redefinition of the spinor by a phase factor, there are essentially two choices for the Majorana-Weyl condition depending on the chirality,
\be
\ba{lll}
\Psi=\pm\Gamma^{(13)}\Psi\,,~~&\mbox{and}&~~
\bar{\Psi}=\Psi^{\dagger}\CA_{+}=\Psi^{t}\CC_{+}\,.
\ea
\label{MW}
\ee 

\subsection{The special unitary Lie superalgebra, ${\mbox{su}(2,2|4)}$} 
Using the twelve dimensional convention, the special unitary Lie superalgebra, ${\mbox{su}(2,2|4)}$, reads simply
\be
\{\CQ,\bar{\CQ}\}=P_{{\13}}\!\left(i\Gamma^{\mu\nu}M_{\mu\nu}-i\Gamma^{ab}M_{ab}\right)\!P_{\13}\,,
\ee
where $\CQ$ satisfies the Majorana-Weyl condition (\ref{MW}) and  $P_{\13}=\half(1\pm\Gamma^{(13)})$.

Explicitly, the sixteen component supercharges, $Q_{\alpha\dalpha}$, carry only the chiral indices for $\mbox{su}(2,2)$ and $\mbox{su}(4)$ so that  the  whole  
 superalgebra, ${\mbox{su}(2,2|4)}$, reads 
\begin{eqnarray}
&&{~~~~~\{Q_{\alpha\dalpha}\,,\bar{Q}^{\beta\dbeta}\}
=i\delta_{\dalpha}{}^{\dbeta}(\rho^{\mu\nu})_{\alpha}{}^{\beta}M_{\mu\nu}
-i\delta_{\alpha}{}^{\beta}(\rho^{ab})_{\dalpha}{}^{\dbeta}M_{ab}\,,}\label{QbarQ}\\
{}\nonumber{}\\
&&{}~~~~~~\{Q_{\alpha\dalpha}\,,Q_{\beta\dbeta}\}=0\,,~~~~~~~~~~~~~~~~~~~~
\{\bar{Q}^{\alpha\dalpha},\,\bar{Q}^{\beta\dbeta}\}=0\,,\label{QQ}\\
{}\nonumber\\
&&{}[M_{\mu\nu},Q_{\alpha\dalpha}]=
(i\textstyle{\frac{1}{2}}\rho_{\mu\nu})_{\alpha}{}^{\beta}Q_{\beta\dalpha}\,,~~~~~~~~
{}[M_{\mu\nu},\bar{Q}^{\alpha\dalpha}]=\bar{Q}^{\beta\dalpha}(
-i\textstyle{\frac{1}{2}}{\rho}_{\mu\nu})_{\beta}{}^{\alpha}\,,\label{M1Q}\\
{}\nonumber{}\\
&&{}[M_{ab},Q_{\alpha\dalpha}]=
(i\textstyle{\frac{1}{2}}\rho_{ab})_{\dalpha}{}^{\dbeta}Q_{\alpha\dbeta}\,,
~~~~~~~~~
{}[M_{ab},\bar{Q}^{\alpha\dalpha}]=\bar{Q}^{\alpha\dbeta}(
-i\textstyle{\frac{1}{2}}{\rho}_{ab})_{\dbeta}{}^{\dalpha}\,,\label{M2Q}\\
{}\nonumber{}\\
&&{}[M_{\mu\nu},M_{\kappa\lambda}]=i(\eta_{\mu\kappa}M_{\nu\lambda}
-\eta_{\mu\lambda}M_{\nu\kappa}
-\eta_{\nu\kappa}M_{\mu\lambda}+\eta_{\nu\lambda}M_{\mu\kappa})\,,\label{MM1}\\
{}\nonumber{}\\
&&{}[M_{ab},M_{cd}]=i(\delta_{ac}M_{bd}-\delta_{ad}M_{bc}
-\delta_{bc}M_{ad}+\delta_{bd}M_{ac})\,,\label{MM2}
\end{eqnarray}
where  $\bar{Q}^{\alpha\dalpha}\equiv A^{\alpha}_{~\beta}(Q^{\dagger})^{\beta\dalpha}$, and    all  the bosonic  generators are hermitian, $(M_{\mu\nu})^{\dagger}=M_{\mu\nu}$, $(M_{ab})^{\dagger}=M_{ab}$.  A few remarks are in order. The relative sign difference   for the $\mbox{so}(2,4)$ and $\mbox{so}(6)$ generators appearing in (\ref{QbarQ}) is crucial for consistency, as required from the Jacobi identity involving  $[Q_{\alpha\dalpha},\{Q_{\beta\dbeta},\bar{Q}^{\gamma\dgamma}\}]$. 
However, the overall sign as well as the chirality choices, namely whether $\rho^{12}\rho^{34}\rho^{56}$ is  $+1$ or $-1$, are solely matter of conventions.\footnote{The freedom for  different chiral choices  reflects two  different  Majorana-Weyl conditions in $2+10$ dimensions, (\ref{MW}).}  
Firstly the over all sign can be flipped by rewriting the superalgebra in terms of the  conjugate supercharges, $(Q^{\prime}=\bar{Q}^{t},\,\bar{Q}^{\prime}=Q^{t}=(Q^{\prime})^{\dagger}A)$ \cite{Kim:2002zg}. 
The equivalence between the different  $\mbox{so}(2,4)$, $\mbox{so}(6)$ chirality choices  becomes clear when we rewrite the superalgebra by the $\mbox{su}(2,2)$, $\mbox{su}(4)$ generators,\footnote{From (\ref{Apm}), $A=A^{\dagger}$.  In fact, as  explained in the next section (\ref{cartan}), one can set $A=\mbox{diag}(-1,-1,+1,+1)$.}
\be
\ba{lll}
T_{\stwo}=-i\textstyle{\frac{1}{4}}\bar{\rho}^{\mu\nu}M_{\mu\nu}\,,~~&~~
T_{\stwo}^{\dagger}=AT_{\stwo}A\,,~~&~~\tr\, T_{\stwo}=0\,,\\
{}&{}&{}\\
T_{\sfour}=-i\textstyle{\frac{1}{4}}\bar{\rho}^{ab}M_{ab}\,,~~&~~
T_{\sfour}^{\dagger}=T_{\sfour}\,,~~&~~\tr\, T_{\sfour}=0\,.
\ea
\label{TM}
\ee
From the completeness relation (\ref{6dfierz}) which does not depend on the chiralities, we get the following  expression for the $\mbox{su}(2,2|4)$ algebra regardless of the chirality choices, 
\begin{eqnarray}
&&
\!\!\!\!\!\!\!\!\!\!\!\!\!\!\!\!\!{} 
[T_{\stwo}{}^{\alpha}{}_{\beta}\,,\,Q_{\gamma\dgamma}]=\delta^{\alpha}_{~\gamma}Q_{\beta\dgamma}
-\textstyle{\frac{1}{4}}\delta^{\alpha}_{~\beta}Q_{\gamma\dgamma}\,,~~\,
{}[T_{\stwo}{}^{\alpha}{}_{\beta}\,,\,T_{\stwo}{}^{\gamma}{}_{\delta}]=
\delta^{\alpha}_{~\delta}T_{\stwo}{}^{\gamma}{}_{\beta}-
\delta^{\gamma}_{~\beta}T_{\stwo}{}^{\alpha}{}_{\delta}\,,\label{su22}\\
&&{}\nonumber\\
&&\!\!\!\!\!\!\!\!\!\!\!\!\!\!\!\!
{}
[T_{\sfour}{}^{\dalpha}{}_{\dbeta}\,,\,Q_{\gamma\dgamma}]
=\delta^{\dalpha}_{~\dgamma}Q_{\gamma\dbeta}
-\textstyle{\frac{1}{4}}\delta^{\dalpha}_{~\dbeta}Q_{\gamma\dgamma}\,,~~
{}[T_{\sfour}{}^{\dalpha}{}_{\dbeta}\,,\,T_{\sfour}{}^{\dgamma}{}_{\ddelta}]=
\delta^{\dalpha}_{~\ddelta}T_{\sfour}{}^{\dgamma}{}_{\dbeta}-
\delta{}^{\dgamma}_{~\dbeta}T_{\sfour}{}^{\dalpha}{}_{\ddelta}\,.\label{su4}
\end{eqnarray}
Essentially the different chiral choices are equivalent to each other up to the  redefinition of  the $\so(2,4)$, $\so(6)$ generators through  (\ref{TM}),  and (\ref{6dfierz}), i.e. $T_{\stwo}=-i\textstyle{\frac{1}{4}}\bar{\rho}^{\mu\nu}M_{\mu\nu}=
-i\textstyle{\frac{1}{4}}\bar{\rho}^{\prime\mu\nu}M^{\prime}_{\mu\nu}$.

\section{$\mbox{u}(1)_{\Y}$ extended  superalgebra and its root structure\label{u1extension}}
Before we proceed further to obtain the noncentral extensions  of the $AdS$ superalgebra, here as an intermediate stage   we  consider the inclusion of  an additional  or ``bonus''  $\mbox{u}(1)_{\Y}$ charge into  the    $\mbox{su}(2,2|4)$ superalgebra which acts as an automorphism of the supergroup.   This $\mbox{u}(1)_{\Y}$ symmetry appears both in the  IIB supergravity and  in the analysis of the four dimensional  $\CN=4$  superconformal group. In IIB supergravity the $\mbox{u}(1)_{\Y}$ symmetry rotates the two  chiral spinors 
(see e.g. \cite{Green:mn}), while on superspace the superconformal group   is  defined in terms of the superspace coordinate transformations  so that the $\mbox{u}(1)_{\Y}$ phase rotation of the odd coordinates  is a part of the superconformal transformations \cite{JHP4D}. However the stringy $\alpha^{\prime}$ correction to the supergravity violates the $\mbox{u}(1)_{\Y}$ symmetry \cite{Intriligator:1998ig,Intriligator:1999ff}, and in $\CN=4$ super Yang-Mills theory  \textit{more than three}-point  correlation functions do not respect the  $\mbox{u}(1)_{\Y}$ symmetry generically \cite{Anselmi:1996dd,Anselmi:1997am,Gonzalez-Rey:1998tk,Eden:1998hh}.  Nevertheless, in our analysis of the   extended  superalgebra, the $\mbox{u}(1)_{\Y}$ charge will always act as an automorphism to the superalgebra either unextended or  noncentrally extended,  so that one can safely  switch it  off    any time.   The main technical advantage  to include the  $\mbox{u}(1)_{\Y}$ charge  is to reduce the number of the  fermionic simple roots from two to one. As the formers involve one chiral as well as one anti-chiral, while the latter corresponds to one chiral only,  the inclusion  will allow us to utilize  the \textit{chirality} of the  superalgebra and simplify the study of the representations of the superalgebra drastically. 

\subsection{Inclusion of a $\mbox{u}(1)_{\Y}$ symmetry\label{u(1)}}
The additional   $\mbox{u}(1)_{\Y}$ charge assigns quantum numbers $+1/2$, $-1/2$ to the supercharges, $Q_{\alpha\dalpha}$, $\bar{Q}^{\alpha\dalpha}$,
\be
\ba{lll}
{}[T_{\uone},Q_{\alpha\dalpha}]=+\half Q_{\alpha\dalpha}\,,~~~&~~~
[T_{\uone},\bar{Q}^{\alpha\dalpha}]=-\half \bar{Q}^{\alpha\dalpha}\,,~~~&~~~
T^{\dagger}_{\uone}=T_{\uone}\,,
\ea\label{u1alge}
\ee
which reflect  the $\mbox{u}(1)_{\Y}$ phase rotation of the chiral spinors.   One of the bosonic subalgebras, $\mbox{su}(4)$, is now extended to $\mbox{u}(4)$,
\begin{equation}
T_{\ufour}{}^{\dalpha}{}_{\dbeta}=T_{\sfour}{}^{\dalpha}{}_{\dbeta}+
\textstyle{\frac{1}{2}}\delta^{\dalpha}_{~\dbeta}\,T_{\uone}\,,
\end{equation}
satisfying
\be
\ba{ll}
[T_{\ufour}{}^{\dalpha}{}_{\dbeta}\,,Q_{\gamma\dgamma}]=
\delta^{\dalpha}_{~\dgamma}Q_{\gamma\dbeta}\,,~~~&~~~
[T_{\ufour}{}^{\dalpha}{}_{\dbeta}\,,\,T_{\ufour}{}^{\dgamma}{}_{\ddelta}]=
\delta^{\dalpha}_{~\ddelta}T_{\ufour}{}^{\dgamma}{}_{\dbeta}
-\delta^{\dgamma}_{~\dbeta}T_{\ufour}{}^{\dalpha}{}_{\ddelta}\,.\label{u4}
\ea
\ee
The additional $\mbox{u}(1)_{\Y}$ charge commutes with all the bosonic generators so that the resulting  superalgebra  is a semi-direct sum of  $\mbox{su}(2,2|4)$ and $\mbox{u}(1)_{\Y}$, or  $\mbox{su}(2,2|4)\,\osplus\,\mbox{u}(1)_{\Y}$.

\subsection{The root structure of $\mbox{su}(2,2|4)\,\osplus\,\mbox{u}(1)_{\Y}$\label{root}}
In this subsection we analyze the root structure of $\mbox{su}(2,2|4)\,\osplus\,\mbox{u}(1)_{\Y}$. Our  analysis is meant to be self-contained and involves much detailed general discussions on the subject. The experienced readers  may skip to the next subsection and only refer to the present one for complements. \newline

We first start with the following $16\times 16$ representation of the bosonic part,
$\mbox{su}(2,2)\oplus\mbox{su}(4)\oplus\mbox{u}(1)_{\Y}$, acting on spinors,
\begin{equation}
\Big(R(M_{\mu\nu})\,,~R(M_{ab}\,)\,,~R(T_{\uone})\Big)
=\Big(-i\textstyle{\frac{1}{2}}\bar{\rho}_{\mu\nu}\otimes
1\,,~1\otimes -i\textstyle{\frac{1}{2}}\bar{\rho}_{ab}\,,~1\otimes \textstyle{\frac{1}{2}}\,\Big)\,,
\end{equation}
which are orthonormal and satisfy the reality condition,
\begin{equation}
\begin{array}{ccc}
\Tr(R_{I}^{\dagger}R_{J})=4\delta_{IJ}\,,~~~&~~~R_{I}^{\dagger}=(A\otimes 1)R_{I}(A\otimes 1)\,,~~~&~~~I,J=1,2,\cdots,31\,.
\end{array}
\label{RI}
\end{equation}
The above representation for $\mbox{su}(2,2)$ is nonunitary. 
This is unavoidable in order to have a finite dimensional representation for 
the noncompact algebra, since any unitary representation of 
a noncompact algebra is always infinite dimensional.\newline

Our choice of the Cartan subalgebra is
\begin{equation}
\H=(T_{\uone},\,M_{12},\,M_{34},\,M_{56},\, M_{78},\,M_{9\,10},\,M_{11\,12})\,.
\end{equation}
Using the $\mbox{SU}(4)$ symmetry, $\rho_{\mu}\rightarrow U\rho_{\mu}U^{T}$, $UU^{\dagger}=1$, which preserves the anti-symmetric property  (\ref{anti-sym}) of $\rho_{\mu}$, 
 we can take the representation of the Cartan subalgebra in a diagonal form. Adopting the bra and ket notations we set
\begin{equation}
\begin{array}{l}
R(M_{12})=\textstyle{\frac{1}{2}}
\left(\,-|1\rangle\langle 1|-|2\rangle\langle 2|+|3\rangle\langle 3|+|4\rangle\langle 4|\,\right)\otimes 1
=\textstyle{\frac{1}{2}}A\otimes 1\,,\\
{}\\
R(M_{34})=\textstyle{\frac{1}{2}}\left(\,-|1\rangle\langle 1|+|2\rangle\langle 2|-|3\rangle\langle 3|+|4\rangle\langle 4|\,\right)\otimes 1\,,\\
{}\\
R(M_{56})=\textstyle{\frac{1}{2}}\left(\,-|1\rangle\langle 1|+|2\rangle\langle 2|+|3\rangle\langle 3|-|4\rangle\langle 4|\,\right)\otimes 1\,,\\
{}\\
R(M_{78})~~\,=1\otimes \textstyle{\frac{1}{2}}\left(\,-|1\rangle\langle 1|-|2\rangle\langle 2|+|3\rangle\langle 3|+|4\rangle\langle
4|\,\right)\,,\\
{}\\
R(M_{9\,10})\,\,=1\otimes \textstyle{\frac{1}{2}}\left(\,-|1\rangle\langle 1|+|2\rangle\langle 2|-|3\rangle\langle 3|+|4\rangle\langle
4|\,\right)\,,\\
{}\\
R(M_{11\,12})=1\otimes \textstyle{\frac{1}{2}}\left(\,-|1\rangle\langle 1|+|2\rangle\langle 2|+|3\rangle\langle 3|-|4\rangle\langle 4|\,\right)\,.
\end{array}
\label{cartan}
\end{equation}

All the  bosonic positive roots and their representations are then given by
\begin{equation}
\begin{array}{ll}
R(\CE_{x})=|2\rangle\langle 1|\otimes 1\,,~~&x=(0,0,1,1,0,0,0)\,,\\
{}&{}\\
R(\CE_{s})=|3\rangle\langle 2|\otimes 1\,,~~&s=(0,1,-1,0,0,0,0)\,,\\
{}&{}\\
R(\CE_{y})=|4\rangle\langle 3|\otimes 1\,,~~&y=(0,0,1,-1,0,0,0)\,,\\
{}&{}\\
R(\CE_{s+x})=|3\rangle\langle 1|\otimes 1\,,~~&s+x=(0,1,0,1,0,0,0)\,,\\
{}&{}\\
R(\CE_{y+s})=|4\rangle\langle 2|\otimes 1\,,~~&y+s=(0,1,0,-1,0,0,0)\,,\\
{}&{}\\
R(\CE_{y+s+x})=|4\rangle\langle 1|\otimes 1\,,~&y+s+x=(0,1,1,0,0,0,0)\,,
\end{array}\label{su2R}
\end{equation}

\begin{equation}
\begin{array}{ll}
R(\CE_{u})=1\otimes |2\rangle\langle 1|\,,~~&u=(0,0,0,0,0,1,1)\,,\\
{}&{}\\
R(\CE_{v})=1\otimes |3\rangle\langle 2|\,,~~&v=(0,0,0,0,1,-1,0)\,,\\
{}&{}\\
R(\CE_{w})=1\otimes |4\rangle\langle 3|\,,~~&w=(0,0,0,0,0,1,-1)\,,\\
{}&{}\\
R(\CE_{v+u})=1\otimes |3\rangle\langle 1|\,,~~&v+u=(0,0,0,0,1,0,1)\,,\\
{}&{}\\
R(\CE_{w+v})=1\otimes |4\rangle\langle 2|\,,~~&w+v=(0,0,0,0,1,0,-1)\,,\\
{}&{}\\
R(\CE_{w+v+u})=1\otimes |4\rangle\langle 1|\,,~&w+v+u=(0,0,0,0,1,1,0)\,,
\end{array}\label{su4R}
\end{equation}
where $x,y,s$ and $u,v,w$ are respectively the $\mbox{su}(2,2)$ and $\mbox{su}(4)$ simple roots. For a given root, $\chi$, the corresponding negative root and its representation  follow  simply from
\be
\ba{lll}
\CE_{-\chi}=\CE_{\chi}^{\dagger}\,,~~~&~~~R(\CE_{-\chi})=(A\otimes 1)\,
R(\CE_{\chi})^{\dagger}\,(A\otimes 1)\,,
\ea
\ee
so that 
\be
R(\CE_{-\chi})=\left\{\ba{ll}
-R(\CE_{\chi})^{\dagger}~~~&~~~\mbox{for~~}\chi\in\{s,\,s+x,\,y+s,\,y+s+x\}\\
{}&{}\\
+R(\CE_{\chi})^{\dagger}~~~&~~~\mbox{otherwise}\ea\right.\,.\label{nR}
\ee
Note that $\{s,\,s+x,\,y+s,\,y+s+x\}$ spans the  noncompact directions of $\su(2,2)$.

Just like $R_{I}$ in (\ref{RI}), $R(\H),\,R(\CE_{+}),\,R(\CE_{-})$ are also orthonormal.  This implies that those two are related by the unitary 
transformation. In particular, the objects appearing in the anti-commutator, $\{\bar{Q},Q\}$, read
\begin{equation}
\begin{array}{ll}
\textstyle{\frac{1}{2}}R(M^{\mu\nu})M_{\mu\nu} =\textstyle{\frac{1}{2}}R(M_{\mu\nu})^{\dagger}M_{\mu\nu} =&R(M_{12})M_{12}+R(M_{34})M_{34}+R(M_{56})M_{56}\\{}&{}\\
{}&\, + \displaystyle{\sum_{\chi\in\Delta_{\stwo}^{+}}}\, \Big(R(\CE_{\chi})^{\dagger}\CE_{\chi}+R(\CE_{-\chi})^{\dagger}\CE_{-\chi}\Big)\,,\\
{}&{}\\
\textstyle{\frac{1}{2}}R(M^{ab})M_{ab}=\textstyle{\frac{1}{2}}R(M_{ab})^{\dagger}M_{ab} =&R(M_{78})M_{78}+R(M_{9\,10})M_{9\,10}+R(M_{11\,12})M_{11\,12}\\{}&{}\\
{}&\, + \displaystyle{\sum_{\chi\in\Delta_{\sfour}^{+}}}\, \Big(R(\CE_{\chi})^{\dagger}\CE_{\chi}+R(\CE_{-\chi})^{\dagger}\CE_{-\chi}\Big)\,,
\end{array}
\label{MR}
\end{equation}
where $\Delta_{\stwo}^{+}$ and $\Delta_{\sfour}^{+}$ denote the sets of all the $\mbox{su}(2,2)$ and $\mbox{su}(4)$ positive roots respectively. 
In fact, for the given set of orthonormal matrices, 
$R(\H),\,R(\CE_{+}),\,R(\CE_{-})$, (\ref{cartan}), (\ref{su2R}), (\ref{su4R}), (\ref{nR}), the formulae above {\textit{define}} all the roots, 
$\CE_{\pm}$, in terms of the hermitian generators, $M_{\mu\nu},M_{ab}$, 
and make sure that $R(\CE_{\pm})$ {\textit{are}} the representations for them. 


In terms of the Cartan subalgebra and  $\mbox{su}(2,2)\oplus\mbox{su}(4)$ roots, $\chi\in\Delta_{\stwo}^{+}\!\cup\,\Delta_{\sfour}^{+}$,  the $\mbox{u}(1)_{\Y}$ extended $AdS_{5}\times S^{5}$ superalgebra, $\mbox{su}(2,2|4)\,\osplus\,\mbox{u}(1)_{\Y}$, reads 
\begin{equation}
\begin{array}{ll}
{}[\H,\CE_{\chi}]=\chi \CE_{\chi}\,,~~~&~~~[\H,\CE_{-\chi}]=-\chi \CE_{-\chi}\,,\\
{}&{}\\
\multicolumn{2}{c}{[\CE_{\chi},\CE_{-\chi}]=\left\{\ba{ll}
-\chi{\cdot\H}~~&~~~\mbox{for~~}\chi\in\{s,\,s+x,\,y+s,\,y+s+x\}\\
{}&{}\\
+\chi{\cdot\H}~~&~~~\mbox{otherwise}\ea\right.\,,}
\end{array}\label{HE}
\end{equation}
~\\

\begin{equation}
\begin{array}{ll}
{}[\CE_{s},\CE_{x}]=\CE_{s+x}\,,~~~&~~~[\CE_{y},\CE_{s}]=\CE_{y+s}\,,\\
{}&{}\\
{}[\CE_{y},\CE_{s+x}]=[\CE_{y+s},\CE_{x}]=\CE_{y+s+x}\,,~~~&~~~[\CE_{x},\CE_{y}]=0\,,\\
{}&{}\\
{}[\CE_{v},\CE_{u}]=\CE_{v+u}\,,~~~&~~~[\CE_{w},\CE_{v}]=\CE_{w+v}\,,\\
{}&{}\\
{}[\CE_{w},\CE_{v+u}]=[\CE_{w+v},\CE_{u}]=\CE_{w+v+u}\,,~~~&~~~[\CE_{u},\CE_{w}]=0\,,
\end{array}
\end{equation}
~\\

\begin{equation}
\begin{array}{ll}
{}[\H,Q_{\alpha\dalpha}]=Q_{\beta\dbeta}R(\H)^{\beta\dbeta}{}_{\alpha\dalpha}\,,~~&~~
[\H,\bar{Q}^{\alpha\dalpha}]=-R(\H)^{\alpha\dalpha}{}_{\beta\dbeta}\bar{Q}^{\beta\dbeta}\,,\\
{}&{}\\
{}[\CE_{\pm\chi},Q_{\alpha\dalpha}]=Q_{\beta\dbeta}
R(\CE_{\pm\chi})^{\beta\dbeta}{}_{\alpha\dalpha}\,, ~~&~~
[\CE_{\pm\chi},\bar{Q}^{\alpha\dalpha}]=
-R(\CE_{\pm\chi})^{\alpha\dalpha}{}_{\beta\dbeta}{}\bar{Q}^{\beta\dbeta}\,,

\end{array}\label{Qalge}
\end{equation}

and

\begin{equation}
\begin{array}{ll}
\{\bar{Q}^{\alpha\dalpha},Q_{\beta\dbeta}\}=&
2\delta^{\dalpha}{}_{\dbeta}\left(\begin{array}{cccc}
f_{1}&\CE_{x}&\CE_{s+x}&\CE_{y+s+x}\\
\CE_{-x}&f_{2}&\CE_{s}&\CE_{y+s}\\
~-\CE_{-s-x}&~-\CE_{-s}&f_{3}&\CE_{y}\\
-\CE_{-y-s-x}&~-\CE_{-y-s}&\CE_{-y}&f_{4}\end{array}\right)^{\!\alpha}_{~~\beta}\\
{}&{}\\
{}&~-2\delta^{\alpha}{}_{\beta}\left(\begin{array}{cccc}
f_{5}&\CE_{u}&\CE_{v+u}&\CE_{w+v+u}\\
\CE_{-u}&f_{6}&\CE_{v}&\CE_{w+v}\\
\CE_{-v-u}&\CE_{-v}&f_{7}&\CE_{w}\\
\CE_{-w-v-u}&\CE_{-w-v}&\CE_{-w}&f_{8}\end{array}\right)^{\!\dalpha}_{~~\dbeta}~~\,,
\end{array}
\label{QQroot}
\end{equation}
where the Cartan subalgebra is organized as 
\begin{equation}
\begin{array}{ll}
f_{1}=\frac{1}{2}(-M_{12}-M_{34}-M_{56})\,,~~&~~f_{2}=\frac{1}{2}(-M_{12}+M_{34}+M_{56})\,,\\
{}&{}\\
f_{3}=\frac{1}{2}(M_{12}-M_{34}+M_{56})\,,~~&~~f_{4}=\frac{1}{2}(M_{12}+M_{34}-M_{56})\,,\\
{}&{}\\
f_{5}=\frac{1}{2}(-M_{78}-M_{9\,10}-M_{11\,12})\,,~~&~~
f_{6}=\frac{1}{2}(-M_{78}+M_{9\,10}+M_{11\,12})\,,\\
{}&{}\\
f_{7}=\frac{1}{2}(M_{78}-M_{9\,10}+M_{11\,12})\,,~~&~~
f_{8}=\frac{1}{2}(M_{78}+M_{9\,10}-M_{11\,12})\,.
\end{array}
\end{equation}
In particular from (\ref{Qalge}), $Q_{11}$ corresponds to the unique fermionic simple root, 
\begin{equation}
\begin{array}{ll}
{}[\H,Q_{11}]=qQ_{11}\,,~~~&~~
q=(\textstyle{+\frac{1}{2},-\frac{1}{2},-\frac{1}{2},-\frac{1}{2},
-\frac{1}{2},-\frac{1}{2},-\frac{1}{2}})\,,\\
{}&{}\\
{}[\CE_{-\chi},Q_{11}]=0~~~&~\mbox{for~all~~}
\chi\in\Delta_{\stwo}^{+}\!\cup\,\Delta_{\sfour}^{+}\,.
\end{array}
\end{equation}
Other fermionic positive roots are $\{q+\chi,\,q+\chi^{\prime},\,q+\chi+\chi^{\prime}\,|\,\chi\in\Delta_{\stwo}^{+},\,
\chi^{\prime}\in\Delta_{\sfour}^{+}\}$.  \newline



The second order Casimir operator, ${\cal C}_{\!{\scriptstyle AdS}}$, reads
\begin{equation}
{\cal C}_{\!{\scriptstyle AdS}}={\cal C}_{\,\stwo}-{\cal C}_{\,\sfour} 
-\textstyle{\frac{1}{2}}Q_{\alpha\dalpha}\bar{Q}^{\alpha\dalpha}\,,
\end{equation}
where ${\cal C}_{\,\stwo}$ and ${\cal C}_{\,\sfour}$ are the $\mbox{su}(2,2)$ and $\mbox{su}(4)$ Casimirs respectively. With the $\su(2,2)$ roots for the noncompact directions, $\Delta^{+}_{s}=\{s,s+x,y+s,y+s+x\}$, they are
\begin{equation}
\begin{array}{ll}
{}{\cal C}_{\,\stwo}&=\textstyle{\frac{1}{2}}M^{\mu\nu}M_{\mu\nu}\\
{}&{}\\
{}&=M_{12}^{2}+M_{34}^{2}+M_{56}^{2}+\{\CE_{x}\,,\,\CE_{-x}\}
+\{\CE_{y}\,,\,\CE_{-y}\}-\!\!\displaystyle{\sum_{\chi\in\Delta^{+}_{s}}}
\{\CE_{\chi}\,,\,\CE_{-\chi}\}\,,\\
{}&{}\\
\multicolumn{2}{l}{{\cal C}_{\,\sfour}=\textstyle{\frac{1}{2}}M^{ab}M_{ab}
=M_{78}^{2}+M_{9\,10}^{2}+M_{11\,12}^{2}+\!\!\displaystyle{\sum_{\chi\in\Delta^{+}_{4}}}
\{\CE_{\chi}\,,\,\CE_{-\chi}\}\,.}
\end{array}\label{Cb}
\end{equation}

\subsection{Nonunitary finite representations\label{rep}}
Starting with an eigenstate of $T_{\uone}$, by acting the negative fermionic roots, $\bar{Q}^{\alpha\dalpha}$, as many as possible - maximally sixteen times surely -  one can obtain a state which is annihilated by all the $\bar{Q}^{\alpha\dalpha}$'s.  Now under the action of the bosonic operators, the state opens up  an irreducible representation
of $\mbox{u}(1)_{\Y}\oplus\mbox{su}(2,2)\oplus\mbox{su}(4)$ or the zeroth floor multiplet. Further from (\ref{Qalge}), any state in the  multiplet  is annihilated by all the fermionic negative roots. 

Generic unitary representations of the noncompact Lie algebra, $\su(2,2)$, are infinite dimensional. However  unitary representations are not of our interest.  In the present paper we focus on  the  nonunitary finite representations  of  $\mbox{su}(2,2|4)\,\osplus\,\mbox{u}(1)_{\Y}$,  denoted by $\R$, satisfying
\begin{equation}
\ba{l}
{}\R_{\H}=(\R_{\H})^{\dagger}\,,\\
{}\\
{}\R_{-\chi}=\left\{\ba{ll}
-(\R_{\chi})^{\dagger}~~~&~~~\mbox{for~~}\chi\in\{s,\,s+x,\,y+s,\,y+s+x\}\\
{}&{}\\
+(\R_{\chi})^{\dagger}~~~&~~~\mbox{otherwise}\ea\right.\,.
\ea\label{forf}
\end{equation}
Namely, just like $R(\CE_{\chi})$ (\ref{nR}), 
the representations of  the roots for  the  $\su(2,2)$ noncompact directions 
are anti-hermitian. This makes the $\su(2,2)$ and $\su(4)$ Casimirs (\ref{Cb}) 
nonnegative definite and ensures finiteness of the representation. 
Essentially,  one can regard $\{\R_{\chi},\,(\R_{\chi})^{\dagger}\,|\, \chi\in\Delta^{+}_{\stwo}\cup\Delta^{+}_{\sfour}\}$  
as a unitary representation of  $\su(4)\oplus\su(4)$, 
since, as an alternative to (\ref{HE}), we have
 \begin{equation}
\begin{array}{lll}
{}[\R_{\H},\R_{\chi}]=\chi \R_{\chi}\,,~~&~~~
[\R_{\H},(\R_{\chi})^{\dagger}]=-\chi (\R_{\chi})^{\dagger}\,,~~&~~~
[\R_{\chi},(\R_{\chi})^{\dagger}]=\chi{\cdot\R_{\H}}\,.
\end{array}
\end{equation}
Consequently for any such irreducible representation there exists a unique \textit{superlowest weight}, $|\Lambda_{L}\rangle$,   being annihilated by all the negative roots,
\begin{equation}
\begin{array}{ll}
\bar{Q}^{\alpha\dalpha}|\Lambda_{L}\rangle=0\,,~~~&~~~\CE_{-\chi}|\Lambda_{L}\rangle=0\,,
~~\chi\in\Delta^{+}_{\stwo}\cup\Delta^{+}_{\sfour}\,.
\end{array}
\end{equation}
The superlowest weight vector  is specified by an arbitrary real number, $r$ and six  non-negative integers or the Dynkin labels, $J_{x},J_{s},J_{y},J_{u},J_{v},J_{w}$,
\begin{equation}
\ba{ll}
\Lambda_{L}=&\Big(\textstyle{r\,,\,
{-\frac{1}{2}(J_{x}+2J_{s}+J_{y})}\,,\,
{-\frac{1}{2}(J_{x}+J_{y})}\,,\,{-\frac{1}{2}(J_{x}-J_{y})}}
\,,\\
{}&{}\\
{}&~~~~~\textstyle{{-\frac{1}{2}(J_{u}+2J_{v}+J_{w})}\,,\,{
-\frac{1}{2}(J_{u}+J_{w})}\,,\,{-\frac{1}{2}(J_{u}-J_{w})}}\Big)\,,
\ea
\label{Lambda}
\end{equation}
satisfying for the $\mbox{su}(2,2)\oplus\mbox{su}(4)$  simple roots, $\chi=x,s,y,u,v,w$ in  (\ref{su2R}) and (\ref{su4R}),
\begin{equation}
\begin{array}{cc}
\displaystyle{-2\frac{\,\chi{\cdot\Lambda_{L}}\,}{\chi^{2}}=J_{\chi}}\,,
~~~&~~~(\CE_{\chi})^{J_{\chi}+1}|\Lambda_{L}\rangle=0\,.
\end{array}
\end{equation}

All the other states are generated by repeated applications of the 
positive roots  on $|\Lambda_{L}\rangle$, and without loss of generality 
one can safely work with the simple roots only, 
$Q_{11}$, $\CE_{\chi},\,\chi=x,s,y,u,v,w$. 
Using the commutator relations,  $[\CE_{\chi},Q]\sim Q$ in (\ref{Qalge}), 
one can always move all the $Q_{11}$'s appearing to either far right or far 
left allowing other fermionic positive roots. 
Therefore the whole supermultiplet is spanned by
\begin{equation}
\CE_{\chi_{m}}\cdots \CE_{\chi_{1}}Q_{\alpha_{n}\dalpha_{n}}\cdots Q_{\alpha_{1}\dalpha_{1}} |\Lambda_{L}\rangle\,, \label{LLQQ}
\end{equation}
or  equivalently
\begin{equation}
Q_{\alpha_{n}\dalpha_{n}}\cdots Q_{\alpha_{1}\dalpha_{1}}
\CE_{\chi_{m}}\cdots \CE_{\chi_{1}}|\Lambda_{L}\rangle\,. \label{finite}
\end{equation}
The latter form makes clear that the whole multiplet is built 
on the zeroth floor by repeated application of the fermionic positive 
roots. As the zeroth floor multiplet has dimension
\cite{Humphreys:dw}
\begin{equation}
\ba{ll}
d_{0}=&\Big[\textstyle{\frac{1}{12}}(J_{x}+1)(J_{s}+1)(J_{y}+1)(J_{x}+J_{s}+2)
(J_{s}+J_{y}+2)(J_{x}+J_{s}+J_{y}+3)\Big]\\
{}&{}\\
{}&~~\times\Big[\textstyle{\frac{1}{12}}(J_{u}+1)(J_{v}+1)(J_{w}+1)
(J_{u}+J_{v}+2)(J_{v}+J_{w}+2)(J_{u}+J_{v}+J_{w}+3)\Big]\,, \ea\label{bdim}
\end{equation}
Eq.(\ref{finite}) implies that the supermultiplet has a finite dimension, $d_{s}$,
\begin{equation}
d_{s}\leq 2^{16}\times d_{0}\,.
\end{equation}

The application of a $Q_{\alpha\dalpha}$ changes the 
$\mbox{u}(1)_{\Y}\oplus\mbox{su}(2,2)\oplus\mbox{su}(4)$ multiplets, 
jumping from one irreducible representation to another. 
In particular, the number of the applied fermionic positive roots determines 
the floor number, zero to sixteen at most. 
Each floor is specified by the $\mbox{u}(1)_{\Y}$ charge,
\begin{equation}
\begin{array}{cc}
r_{N}=r+\half N\,,~~~&~~~N=0,1,2,\cdots,16.
\end{array}
\end{equation}

Each of the zeroth and the highest floors  forms an irreducible representation 
of $\mbox{u}(1)_{\Y}\oplus\mbox{su}(2,2)\oplus\mbox{su}(4)$, while
other floors are in general reducible and decompose  into irreducible ones.  
All the irreducible representations for 
$\mbox{u}(1)_{\Y}\oplus\mbox{su}(2,2)\oplus\mbox{su}(4)$ are 
specified by their own lowest weights, $\lambda_{L}$, 
annihilated by all the bosonic negative roots,
\begin{equation}
\ba{ll}
\lambda_{L}=&\Big(\textstyle{r+\half N\,,\,
{-\frac{1}{2}(j_{x}+2j_{s}+j_{y})}\,,\,
{-\frac{1}{2}(j_{x}+j_{y})}\,,\,{-\frac{1}{2}(j_{x}-j_{y})}}
\,,\\
{}&{}\\
{}&~~~~~\textstyle{{-\frac{1}{2}(j_{u}+2j_{v}+j_{w})}\,,\,{
-\frac{1}{2}(j_{u}+j_{w})}\,,\,{-\frac{1}{2}(j_{u}-j_{w})}}\Big)\,.
\ea
\end{equation}
The corresponding highest weight is then \cite{Fulton}
\begin{equation}
\ba{ll}
\lambda_{H}=&\Big(\textstyle{r+\half N\,,\,
{{\frac{1}{2}(j_{y}+2j_{s}+j_{x})}\,,\,{
\frac{1}{2}(j_{y}+j_{x})}\,,\,{\frac{1}{2}(j_{y}-j_{x})}}}\,,\\
{}&{}\\
{}&~~~~~\textstyle{{{\frac{1}{2}(j_{w}+2j_{v}+j_{u})}\,,\,{
\frac{1}{2}(j_{w}+j_{u})}\,,\,{\frac{1}{2}(j_{w}-j_{u})}}}\Big)\,,
\ea
\end{equation}
while the dimension is given by (\ref{bdim}) with $J\leftrightarrow j$.\\

In general, different  orderings in the multiplications of the positive roots 
on the superlowest weight may result in degeneracy for states of the 
same weight vector. To verify the possible degeneracy one should check whether 
a state can be rewritten as the other through changes of orderings using 
the super-commutation relations of the superalgebra. Especially for  irreducible 
representations, if a state is annihilated by all the negative simple roots 
- hence by all the negative roots - the state must be either the 
superlowest weight or trivial. This provides an  alternative  criteria 
to distinguish or identify any given two states of the same weight vector 
in a representation.\\

The particular representation we have in mind for the noncentral extension of $\su(2,2|4)$ superalgebra to be carried out in the next section   is a kind of   \textit{adjoint representation} where $\mbox{su}(2,2|4)$ generators act in the adjoint manner on brane charges which carry  finite number of $\mbox{su}(2,2)\oplus\mbox{su}(4)$ spinor indices, e.g. $Z_{\alpha_{1}\cdots\alpha_{k}\dalpha_{1}\cdots\dalpha_{l}}
{}^{\beta_{1}\cdots\beta_{m}\dbeta_{1}\cdots\dbeta_{n}}$.  Naturally the dimension of the representation  is finite and $\R_{\H}$, $\R_{\pm\chi}$ satisfy the condition (\ref{forf}), since  they are essentially given by $R(\H)$, $R(\CE_{\pm\chi})$,  $-R(\H)^{t}$, $-R(\CE_{\pm\chi})^{t}$, depending on whether the spinor indices are  lower or upper ones. 
Acting the fermionic positive roots, $Q_{\alpha\dalpha}$,  on the ground floor as in (\ref{finite}), all possible states in the supermultiplet are built up, which in fact by definition gives representations of the  fermionic positive roots,  $\R_{\alpha\dalpha}$. 
On the other hand, the representations of the fermionic negative roots, 
$\R^{\alpha\dalpha}$, should be read off from explicit manipulation 
of their actions on all the existing states  
utilizing the  anti-commutation relation of the superalgebra 
until it hits the superlowest weight to terminate the procedure. 
For the adjoint representation we have\footnote{Note that if we 
assumed $\R^{\alpha\dalpha}=\R_{\alpha\dalpha}^{\dagger}$, 
then our  representation would coincide with the unitary representation of 
the $\su(4|4)$ superalgebra. 
However, then, from $\{\R^{\alpha\dalpha},\R_{\alpha\dalpha}\}=0$ and 
its positive definite property, the representation should have been trivial. 
In fact, the precise relation of $\R^{\alpha\dalpha}$ to 
$\R_{\alpha\dalpha}^{\dagger}$ can be obtained only when we complete 
the vector space of the representation  by the  complex conjugate.}

\begin{equation}
\begin{array}{ll}
\{\R^{\alpha\dalpha},\R_{\beta\dbeta}\}=&
2\delta^{\dalpha}{}_{\dbeta}\left(\begin{array}{cccc}
\R_{{1}}&\R_{x}&\R_{s+x}&\R_{y+s+x}\\
\R_{x}^{\dagger}&\R_{{2}}&\R_{s}&\R_{y+s}\\
~\R_{s+x}^{\dagger}&~\R_{s}^{\dagger}&\R_{{3}}&\R_{y}\\
\R_{y+s+x}^{\dagger}&~\R_{y+s}^{\dagger}&\R_{y}^{\dagger}
&\R_{{4}}\end{array}\right)^{\!\alpha}_{~~\beta}\\
{}&{}\\
{}&~-2\delta^{\alpha}{}_{\beta}\left(\begin{array}{cccc}
\R_{{5}}&\R_{u}&\R_{v+u}&\R_{w+v+u}\\
\R_{u}^{\dagger}&\R_{{6}}&\R_{v}&\R_{w+v}\\
\R_{v+u}^{\dagger}&\R_{v}^{\dagger}&\R_{{7}}&\R_{w}\\
\R_{w+v+u}^{\dagger}&\R_{w+v}^{\dagger}&
\R_{w}^{\dagger}&\R_{{8}}\end{array}\right)^{\!\dalpha}_{~~\dbeta}~~\,.
\end{array}
\end{equation}
Note that, compared to (\ref{QQroot}), there is no minus sign for the generators of the noncompact directions in $\su(2,2)$.\newline

As usual, for some small irreducible representations  of the $\su(4)$ algebra, 
we may denote them simply by their dimensions, instead of the Dynkin labels,
\be
\ba{llll}
4\sim(1,0,0)\,,~~~&~~\bar{4}\sim(0,0,1)\,,~~~&~~6\sim(0,1,0)\,,~~~&~~10\sim(2,0,0)\,,\\
{}&{}&{}&{}\\
15\sim(1,0,1)\,,~~~&~~20\sim(3,0,0)\,,~~~&~~20^{\prime}\sim(1,1,0)\,,~~~&~~ 20^{\prime\prime}\sim(0,2,0)\,,\\
{}&{}&{}&{}\\
 35\sim(4,0,0)\,,~~~&~~36\sim(2,0,1)\,,~~~&~~45\sim(2,1,0)\,,~~~&~~60\sim(1,2,0)\,,\\
 {}&{}&{}&{}\\
 70\sim(1,0,3)\,,~~~&~~84\sim(3,1,0)\,.&{}&{}
 \ea
 \ee

\newpage


\section{Noncentral extensions of the  $AdS_{5}\times S^{5}$ superalgebra \label{main}}
One possible way to obtain the noncentral extension of the $\mbox{su}(2,2|4)$ superalgebra is to perform the Witten-Olive type analysis on the four dimensional $\CN=4$ super Yang-Mills theory~\cite{Witten:mh,Osborn:tq}.  Namely starting with the explicit expressions for the supercharges,  including the special superconformal charges too, one may evaluate  the anti-commutators of them to see what kinds of surface  terms appear. In principle, one gets 
\be
\ba{l}
{}\{\bar{Q}^{\alpha\dalpha},Q_{\beta\dbeta}\}=
4\delta^{\dalpha}_{~\dbeta}T_{\stwo}{}^{\alpha}{}_{\beta}-4
\delta^{\alpha}_{~\beta}T_{\sfour}{}^{\dalpha}{}_{\dbeta}+
{\RL}^{\alpha\dalpha}{}_{\beta\dbeta}\,,\\
{}\\
{}\{Q_{\alpha\dalpha}\,,\,Q_{\beta\dbeta}\}=\CZ_{\alpha\dalpha\beta\dbeta}\,.\label{QQG}
\ea
\ee
Here ${\RL}^{\alpha\dalpha}{}_{\beta\dbeta}$ and $\CZ_{\alpha\dalpha\beta\dbeta}$ correspond to the possible surface integrals or the brane charges, and they can further decompose into 
$(15,15)\oplus(1,15)\oplus(15,1)\oplus(1,1)$ and $(6,6)\oplus(10,10)$,
\be
\ba{l}
{\RL}^{\alpha\dalpha}{}_{\beta\dbeta}=\sR^{\alpha\dalpha}{}_{\beta\dbeta}+
\textstyle{\frac{1}{4}}\delta^{\dalpha}{}_{\dbeta}{\sR}^{\alpha}{}_{\beta}
-\textstyle{\frac{1}{4}}\delta^{\alpha}{}_{\beta}{\sR}^{\dalpha}{}_{\dbeta}
+\textstyle{\frac{1}{16}}\delta^{\alpha}{}_{\beta}\delta^{\dalpha}{}_{\dbeta}\sR\,,\\
{}\\
\CZ_{\alpha\dalpha\beta\dbeta}=
Z^{{\scriptscriptstyle{(6,6)}}}_{\alpha\beta\dalpha\dbeta}+
Z^{{\scriptscriptstyle{(10,10)}}}_{\alpha\beta\dalpha\dbeta}\,,
\ea
\ee
satisfying the traceless and  symmetric  properties, 
\be
\ba{ll}
{\sR}^{\alpha}{}_{\alpha}=0\,,~~~~{\sR}^{\dalpha}{}_{\dalpha}=0\,,~~&~~
\sR^{\alpha\dalpha}{}_{\alpha\dbeta}=0\,,~~~~\sR^{\alpha\dalpha}{}_{\beta\dalpha}=0\,,\\
{}&{}\\
Z^{{\scriptscriptstyle{(6,6)}}}_{\alpha\beta\dalpha\dbeta}=
Z^{{\scriptscriptstyle{(6,6)}}}_{[\alpha\beta][\dalpha\dbeta]}\,,~~&~~
Z^{{\scriptscriptstyle{(10,10)}}}_{\alpha\beta\dalpha\dbeta}=
Z^{{\scriptscriptstyle{(10,10)}}}_{(\alpha\beta)(\dalpha\dbeta)}\,.
\ea
\ee

Using the $4\times4$ matrices, $\rho^{\mu}, \rho^{a}$,  they can be rewritten as
\begin{equation}
\ba{cl}
{\sR}^{\alpha}{}_{\beta}=-i\half
(\bar{\rho}^{\mu\nu})^{\alpha}{}_{\beta}{\sR}_{\mu\nu}\,,
~~
{\sR}^{\dalpha}{}_{\dbeta}=-i\half
(\bar{\rho}^{ab})^{\dalpha}{}_{\dbeta}{\sR}_{ab}\,,\,&
\sR^{\alpha\dalpha}{}_{\beta\dbeta}=\textstyle{\frac{1}{4}}
(\bar{\rho}^{\mu\nu})^{\alpha}{}_{\beta}
(\bar{\rho}^{ab})^{\dalpha}{}_{\dbeta}\sR_{\mu\nu ab}\,,\\
{}&{}\\
Z^{{\scriptscriptstyle{(6,6)}}}_{\alpha\beta\dalpha\dbeta}=
(\rho^{\mu})_{\alpha\beta}(\rho^{a})_{\dalpha\dbeta}Z_{\mu a}\,,\,&
Z^{{\scriptscriptstyle{(10,10)}}}_{\alpha\beta\dalpha\dbeta}=
\textstyle{\frac{1}{144}}(\rho^{\mu\nu\lambda})_{\alpha\beta}(\rho^{abc})_{\dalpha\dbeta}
Z^{-}_{\mu\nu\lambda abc}\,,
\ea
\end{equation}
where  $\sR_{\mu\nu ab},\,{\sR}_{\mu\nu},\,{\sR}_{ab},\,\sR$ are all hermitian, and   from (\ref{iden}), $Z^{-}_{\mu\nu\lambda abc}$ is anti-self-dual for each $\so(2,4)$ and $\so(6)$ indices,
\be
Z^{-}_{\mu\nu\lambda abc}=-i\textstyle{\frac{1}{6}}
\epsilon_{\mu\nu\lambda}{}^{\kappa\sigma\tau}Z^{-}_{\kappa\sigma\tau abc}
=-i\textstyle{\frac{1}{6}}
\epsilon_{abc}{}^{def}Z^{-}_{\mu\nu\lambda def}\,.
\ee
Physically, $\sR_{\mu\nu ab}$, $Z_{\mu a}$, $Z^{-}_{\mu\nu\lambda abc}$ 
correspond to the $D3$, ${F1/D1}$, ${D5/N\!S}5$ branes. 
A simple way to see this  is to begin with a single probe brane 
orthogonally intersecting with a stack of $D3$ branes in flat space 
and to take the `near horizon limit' for the $D3$ branes  in the sense of 
Maldacena's original approach to the $AdS$/CFT correspondence 
\cite{Maldacena:1997re}.
The brane configurations preserve eight supercharges in flat space,  
which is enhanced to sixteen in the AdS limit,  
as they have four Neumann-Dirichlet  directions for  the $D$-branes  ($N\!S$ branes are related by S-duality). More specifically, a brane charge with $p$ indices for  $\mbox{so}(2,4)$ and $q$ indices for $\mbox{so}(6)$ corresponds to the $(p+q-1)$ brane wrapping an $AdS_{p+1} \times S^{q-1}$ subspace maximally embedded in 
$AdS_{5} \times S^{5}$ \cite{Skenderis:2002vf}.
This result implies that there is no brane configuration corresponding to 
$H_{\mu\nu},\,H_{ab},\,H$ charges. 

The brane analysis also agrees with the field theory  result obtained by  Osborn~\cite{Osborn:tq} who  showed that there appears no $H_{\mu\nu},\,H_{ab},\,H$ terms  in the expression of the anti-commutator between  the two ordinary  supercharges. This indicates  that, at least, some components of $\sR_{\mu\nu},\sR_{ab}$ are identically  vanishing in the extended superalgebra. Then the covariance under the  $\so(2,4)\oplus\so(6)$ rotation makes sure that all of them are indeed absent. Hence we conclude
\be
\ba{llll}
\sR_{\mu\nu}=0\,,~~&~~\sR_{ab}=0\,,~~&~~\sR=0\,,~~&~~
\RL^{\alpha\dalpha}{}_{\beta\dbeta}=\sR^{\alpha\dalpha}{}_{\beta\dbeta}\,.
\ea
\ee

As noted by  Peeters and Zamaklar~\cite{Peeters:2003vz}, due to the Jacobi identity involving $Q$, $\bar{Q}$ and a  brane charge, the commutators between the brane charges and supercharges should not vanish, e.g. $[\sR_{\mu\nu ab}\,,\,Q_{\alpha\dalpha}]\neq 0$ if 
$\sR_{\mu\nu ab}\neq 0$. Naturally this leads to a noncentral extension of the superalgebra, $\su(2,2|4)$. In the rest of the present paper, we study the noncentral extension in a group theoretical manner, rather than pursuing the Witten-Olive type analysis on the four dimensional $\CN=4$ super Yang-Mills theory.

\subsection{Generic features of the  extended superalgebra} 
In our terminology, \textit{brane charges} are, by definition, the space integrals of the total derivative terms or the surface integrals. In particular, they are not symmetry generators of the corresponding field theory, and hence  they are not forbidden by  the Coleman-Mandula theorem~\cite{Coleman:ad}.  Some  immediate important  consequences are as follows.  Firstly \textbf{the  super-commutator involving a brane charge is also a brane charge}, since whatever comes out  should remain as a surface integral.   
Furthermore, \textbf{all the brane charges super-commute with each other}, 
since one can take the two radii of the spatial infinite spheres, $S^{\,2}$, to be finitely different so that the two surfaces have no contact point.\footnote{One  exceptional   case is the square of a fermionic brane charge, which diverges in general.  Either we can  take again  two different radii at spatial infinities and set  it vanish as a kind of regularization scheme, or leave them undetermined. In any case, our main results are not affected by this subtlety.} As a result all the bosonic brane charges can be diagonalized  simultaneously and provide good quantum numbers.  Schematically we have\footnote{The super-commutator is defined to be $[\CO_{\scriptstyle{A}}\,,\,\CO_{\scriptstyle{B}}\}=
\CO_{\scriptstyle{A}}\CO_{\scriptstyle{B}}-(-1)^{\#_{\scriptstyle{A}}\#_{\scriptstyle{B}}}
\CO_{\scriptstyle{B}}\CO_{\scriptstyle{A}}$, where $\#_{\scriptstyle{A}}$ is zero or one depending whether $\CO_{\scriptstyle{A}}$ is bosonic or fermionic.}
\be
\ba{l}
{}[\CO_{\scriptstyle{A}}\,,\,\CO_{\scriptstyle{B}}\}
=c_{{\scriptstyle{AB}}}{}^{{\scriptstyle{C}}}
\CO_{{\scriptstyle{C}}}\,+
\,d_{{\scriptstyle{AB}}}{}^{{\scriptstyle{I}}}\B_{\scriptstyle{I}}\,,\\
{}\\
{}[\B_{\scriptstyle{I}}\,,\,\B_{\scriptstyle{J}}\}=0\,,\\
{}\\
{}[\CO_{\scriptstyle{A}}\,,\,\B_{\scriptstyle{I}}\}
=f_{{\scriptstyle{AI}}}{}^{{\scriptstyle{J}}}\B_{\scriptstyle{J}}\,,
\ea
\ee
where $\CO_{\scriptstyle{A}}$ denotes  the old generators in the unextended superalgebra, $\su(2,2|4)$, with the  structure constant, $c_{{\scriptstyle{AB}}}{}^{{\scriptstyle{C}}}$, while $\B_{\scriptstyle{I}}$ corresponds  to   the brane charges. 

For  consistency, it is necessary and sufficient to require  the extended superalgebra to satisfy the Jacobi identity, as the structure constants which are ordinary $c$-numbers will then realize \textit{a}  representation or the adjoint representation.  In our case, the Jacobi identities involving more than one brane charges are trivial so that  there exist essentially two types of Jacobi identities to consider :
\begin{eqnarray}
&&{}[\CO_{\scriptstyle{A}}\,,[\CO_{\scriptstyle{B}}\,,\B_{\scriptstyle{I}}\}\}
-(-1)^{\#_{\scriptstyle{A}}\#_{\scriptstyle{B}}}
[\CO_{\scriptstyle{B}}\,,[\CO_{\scriptstyle{A}}\,,\B_{\scriptstyle{I}}\}\}
=[[\CO_{\scriptstyle{A}}\,,\CO_{\scriptstyle{B}}\}\,,\B_{\scriptstyle{I}}\}\,,\label{adjoint}\\
{}\nonumber\\
&&{}[\CO_{\scriptstyle{A}}\,,[\CO_{\scriptstyle{B}}\,,\CO_{\scriptstyle{C}}\}\}
-(-1)^{\#_{\scriptstyle{A}}\#_{\scriptstyle{B}}}
[\CO_{\scriptstyle{B}}\,,[\CO_{\scriptstyle{A}}\,,\CO_{\scriptstyle{C}}\}\}
=[[\CO_{\scriptstyle{A}}\,,\CO_{\scriptstyle{B}}\}\,,\CO_{\scriptstyle{C}}\}\,.\label{constraint}
\end{eqnarray}
The first identity clearly shows that \textbf{the brane charges form a representation realized by the adjoint actions of  the generators in the original unextended superalgebra}, while the second one indicates that  \textbf{the adjoint representation is subject to some constraints}. In particular, the dimension of the adjoint representation is finite, meaning that there are only finitely many brane charges.

Requiring  that the brane charges transform covariantly for the  ${\su(2,2)\oplus\su(4)}$ generators, as described in subsection (\ref{rep}), any Jacobi identity involving the ${\su(2,2)\oplus\su(4)}$ generators holds automatically. Therefore the only nontrivial constraints come from Jacobi identities containing either three $Q$'s ~or~ two $Q$'s and one $\bar{Q}$,
\bea
&&{}[Q_{\alpha\dalpha}\,,\,\CZ_{\beta\dbeta\gamma\dgamma}]+
[Q_{\beta\dbeta},\CZ_{\gamma\dgamma\alpha\dalpha}]+
[Q_{\gamma\dgamma}\,,\,\CZ_{\alpha\dalpha\beta\dbeta}]\equiv \Psi_{\alpha\beta\gamma\dalpha\dbeta\dgamma}=0\,,\label{remain}\\
{}\nonumber\\
&&{}[Q_{\alpha\dalpha}\,,\,{\sR}^{\gamma\dgamma}{}_{\beta\dbeta}]+
[Q_{\beta\dbeta}\,,\,{\sR}^{\gamma\dgamma}{}_{\alpha\dalpha}]+
[\bar{Q}^{\gamma\dgamma}\,,\,\CZ_{\alpha\dalpha\beta\dbeta}]=0\,.\label{superlowest}
\eea
To obtain the  extended superalgebra, one needs to look for  adjoint representations of the original unextended superalgebra such that it  contains $\CZ_{\alpha\dalpha\beta\dbeta}$, ${\sR}^{\alpha\dalpha}{}_{\beta\dbeta}$ and satisfies the constraints above. However, this  group theoretically well defined  problem does not lead to a unique solution, essentially because the relevant superlowest weights are not specified yet, and due to the  nonunitary property of the adjoint representation, the states which can decouple may not decouple. In fact, we expect the ``correctly" extended superalgebra, which can be in principle uniquely obtained from the Witten-Olive type analysis on the  $\CN=4$ super Yang-Mills theory,   leads to a   \textit{reducible} adjoint representation for the brane charges, containing    more than one irreducible supermultiplets. The physical reason is that the $D1$, $D3$, $D5$ branes should be able to exist separately,  not necessarily weaved by one another.

The filtering of the reducible  representation into each irreducible one can be done   by restricting the full  Hilbert space of the Yang-Mills theory in a suitable way, and this will enable us  to obtain the physically relevant   noncentral extensions.

Firstly we raise the question, `what is the relevance of the strictly unextended superalgebra, $\su(2,2|4)$, to the Yang-Mills theory, if the ``correct" superalgebra of the theory is an extended one not the unextended one?'  The answer is simple. Consider a subspace of  the full  Hilbert space which is annihilated by all the brane charges. Clearly such a subspace forms an invariant subspace for the extended  superalgebra,  and on the subspace the brane charges have the trivial representations.  In other words, the unextended superalgebra is only for the elementary particles in the theory not for the branes, as one can naturally expect.

Now we consider a less restricted subspace of the full Hilbert space.  Namely, we focus on the subspace, $V$, which satisfies the following two properties. First it is annihilated by   the $D3$ brane charges, ${\sR}_{\mu\nu ab}$, and second it is invariant under the action of all the supercharges,
\begin{equation}
\ba{lll}
\sR_{\mu\nu ab}V=0\,,~~~&~~~Q_{\alpha\dalpha}V\subset V\,,~~~&~~~\bar{Q}^{\alpha\dalpha}V\subset V\,.
\ea
\ee
It follows that $V$ is in fact an invariant subspace for the fully extended superalgebra, since all other generators can be constructed from the supercharges. Furthermore, we get
\be
[Q_{\alpha\dalpha}\,,\,{\sR}^{\gamma\dgamma}{}_{\beta\dbeta}]\,V=0\,.
\ee
Clearly on the  subspace, $V$,  the representations of $\,{\sR}^{\alpha\dalpha}{}_{\beta\dbeta}$ and $ [Q_{\alpha\dalpha}\,,\,{\sR}^{\gamma\dgamma}{}_{\beta\dbeta}]$ are   trivial,  
and   Eq.(\ref{superlowest}) gets simplified to show that $\CZ_{\alpha\dalpha\beta\dbeta}$ forms the ground floors of the adjoint representations we are looking for, 
\be
[\,\bar{Q}^{\gamma\dgamma}\,,\,\CZ_{\alpha\dalpha\beta\dbeta}\,]=0\,.\label{zeroth}
\ee
Moreover, as it decomposes into $(6,6)$ and $(10,10)$,  there exist  two superlowest weights, and hence two irreducible adjoint  representations.  They can be treated separately, and  we only need to impose the remaining constraint, $\Psi_{\alpha\beta\gamma\dalpha\dbeta\dgamma}=0$, (\ref{remain}).  \newline

Direct calculation, using  (\ref{QQG}) and (\ref{zeroth}) only, shows that 
\be
\{\,\bar{Q}^{\kappa\dkappa},\Psi_{\alpha\beta\gamma\dalpha\dbeta\dgamma}\,\}=0\,,~~~~~~~~
\mbox{identically}\,.
\ee
Surely this is a necessary condition for the consistent decoupling of $\Psi_{\alpha\beta\gamma\dalpha\dbeta\dgamma}$ in the adjoint representation.

\subsection{Electro-magnetic  extension\label{mainsub}} 
The aim of the present subsection is to obtain   the  noncentral extension of the  superalgebra, $\su(2,2|4)$,  which contains the $F1/D1$ or the electro-magnetic  charge,  $Z_{\mu a}$, in the anti-commutator of the supercharges, 
\be
\ba{ll}
{}\{Q_{\alpha\dalpha}\,,\,Q_{\beta\dbeta}\}=\textstyle{\frac{1}{4}}
\epsilon_{\alpha\beta\gamma\delta}\epsilon_{\dalpha\dbeta\dgamma\ddelta}
\B^{\gamma\delta\dgamma\ddelta}\,,~~~&~~\{\bar{Q}^{\alpha\dalpha},Q_{\beta\dbeta}\}=
4\delta^{\dalpha}_{~\dbeta}T_{\stwo}{}^{\alpha}{}_{\beta}-4
\delta^{\alpha}_{~\beta}T_{\sfour}{}^{\dalpha}{}_{\dbeta}\,,
\ea\label{D1D3sa}
\ee
where, for the later convenience, we have  raised the spinor indices of  the electro-magnetic  charge by the totally anti-symmetric four form tensors, 
\be
\B^{\alpha\beta\dalpha\dbeta}=\textstyle{\frac{1}{4}}\epsilon^{\alpha\beta\gamma\delta} \epsilon^{\dalpha\dbeta\dgamma\ddelta}
Z^{{\scriptscriptstyle{(6,6)}}}_{\gamma\delta\dgamma\ddelta}=
(\bar{\rho}^{\mu})^{\alpha\beta}(\bar{\rho}^{a})^{\dalpha\dbeta}Z_{\mu a}\,.
\ee
As the brane charge, $\B^{\alpha\beta\dalpha\dbeta}$, can not be central,   the superalgebra, $\su(2,2|4)$, gets a noncentral extension inevitably.  The extension will be  uniquely determined, and   the corresponding extended superalgebra  can be regarded as the superalgebra  of the  $\CN=4$ super Yang-Mills theory restricted on  the  `$D3$, $D5$ free'  \,Hilbert space or    ${H}^{\alpha\dalpha}{}_{\beta\dbeta}\equiv0$, $Z^{{\scriptscriptstyle{(10,10)}}}_{\alpha\beta\dalpha\dbeta}\equiv 0$.\newline

From the decomposition of  the tensor product,
\be
(4,4)\otimes(6,6)=(20,20)\oplus(\bar{4},20)\oplus(20,\bar{4})\oplus(\bar{4},\bar{4})\,,
\ee
we write,  for the  first floor of the adjoint representation, 
\be
\ba{ll}
{}[Q_{\alpha\dalpha},\B^{\beta\gamma\dbeta\dgamma}]=& N_{\alpha\dalpha}{}^{\beta\gamma\dbeta\dgamma}
+\textstyle{\frac{1}{9}}\delta_{\alpha}^{\,\beta}\delta_{\dalpha}^{\,\dbeta}N^{\gamma\dgamma}
-\textstyle{\frac{1}{9}}\delta_{\alpha}^{\,\beta}\delta_{\dalpha}^{\,\dgamma}N^{\gamma\dbeta}
-\textstyle{\frac{1}{9}}\delta_{\alpha}^{\,\gamma}\delta_{\dalpha}^{\,\dbeta}N^{\beta\dgamma}
+\textstyle{\frac{1}{9}}\delta_{\alpha}^{\,\gamma}\delta_{\dalpha}^{\,\dgamma}N^{\beta\dbeta}\\
{}&{}\\
{}&-\textstyle{\frac{1}{3}}\delta_{\alpha}^{\,\beta}\B^{\gamma}{}_{\dalpha}{}^{\dbeta\dgamma}+
\textstyle{\frac{1}{3}}\delta_{\alpha}^{\,\gamma}\B^{\beta}{}_{\dalpha}{}^{\dbeta\dgamma}
-\textstyle{\frac{1}{3}}\delta_{\dalpha}^{\,\dbeta}\B^{\beta\gamma\dgamma}{}_{\alpha}+
\textstyle{\frac{1}{3}}\delta_{\dalpha}^{\,\dgamma}\B^{\beta\gamma\dbeta}{}_{\alpha}\,,
\ea\label{QBpre}
\ee
where each tensor belongs to different $\su(2,2)\oplus\su(4)$ irreducible representation as  they are traceless and anti-symmetric, 
\be
\ba{lll}
N_{\alpha\dalpha}{}^{\alpha\gamma\dbeta\dgamma}=0\,,~~~&~~~
N_{\alpha\dalpha}{}^{\beta\gamma\dbeta\dgamma}=
N_{\alpha\dalpha}{}^{[\beta\gamma][\dbeta\dgamma]}\,,~~~&:~
(20,20)\,,\\
{}&{}&{}\\
B^{\gamma}{}_{\dalpha}{}^{\dalpha\dgamma}=0\,,~~~&~~~
B^{\gamma}{}_{\dalpha}{}^{\dbeta\dgamma}=B^{\gamma}{}_{\dalpha}{}^{[\dbeta\dgamma]}\,,~~~&:~
(\bar{4},20)\,,\\
{}&{}&{}\\
\B^{\alpha\gamma\dgamma}{}_{\alpha}=0\,,~~~&~~~
\B^{\beta\gamma\dgamma}{}_{\alpha}=\B^{[\beta\gamma]\dgamma}{}_{\alpha}\,,~~~&:~(20,\bar{4})\,.
\ea
\ee
In terms of the decomposition, the six form tensor reads
\be
\Psi_{\alpha\beta\gamma\dalpha\dbeta\dgamma}=\textstyle{\frac{1}{4}}
\epsilon_{\alpha\beta\rho\varepsilon}\epsilon_{\dalpha\dbeta\drho\dvarepsilon}
N_{\gamma\dgamma}{}^{\rho\varepsilon\drho\dvarepsilon}+\textstyle{\frac{1}{4}}
\epsilon_{\beta\gamma\rho\varepsilon}\epsilon_{\dbeta\dgamma\drho\dvarepsilon}
N_{\alpha\dalpha}{}^{\rho\varepsilon\drho\dvarepsilon}+\textstyle{\frac{1}{4}}
\epsilon_{\gamma\alpha\rho\varepsilon}\epsilon_{\dgamma\dalpha\drho\dvarepsilon}
N_{\beta\dbeta}{}^{\rho\varepsilon\drho\dvarepsilon}
+\textstyle{\frac{1}{3}}
\epsilon_{\alpha\beta\gamma\varepsilon}\epsilon_{\dalpha\dbeta\dgamma\dvarepsilon}
N^{\varepsilon\dvarepsilon}\,.
\ee
In particular, 
\be
N^{\alpha\dalpha}=\textstyle{\frac{1}{12}}\epsilon^{\alpha\beta\gamma\delta} \epsilon^{\dalpha\dbeta\dgamma\ddelta}\Psi_{\beta\gamma\delta\dbeta\dgamma\ddelta}\,.
\ee
Hence the constraint, $\Psi\equiv 0$, is equivalent to 
\be
\ba{ll} 
N_{\alpha\dalpha}{}^{\beta\gamma\dbeta\dgamma}\equiv0\,,~~~~&~~~N^{\alpha\dalpha}\equiv0\,,
\ea
\label{Nconstraint}
\ee
which imply  only the $(\bar{4},20)$ and $(20,\bar{4})$ tensors survive and others decouple.

Consequently the commutation relation for the first floor, (\ref{QBpre}),  becomes simplified, and  other higher floors can be constructed recurrently. It turns out that the construction terminates on the fourth floor, and the resulting adjoint representation is  of the following unique form,\newline
\newline
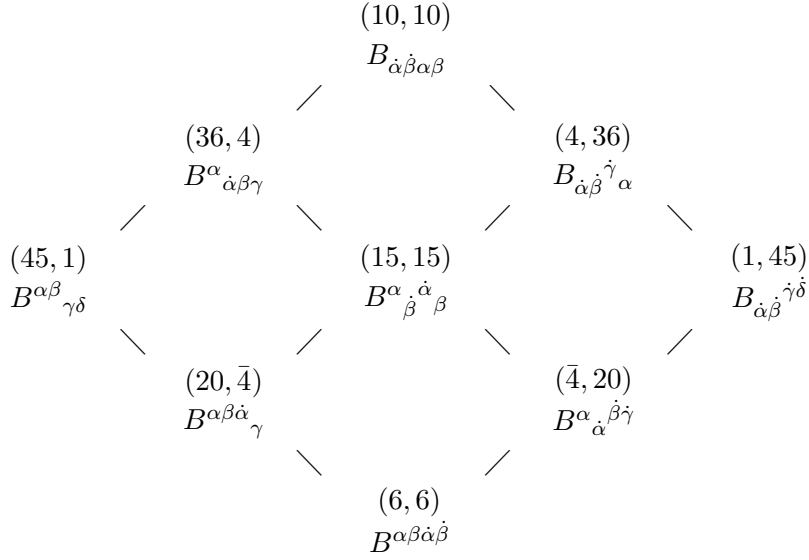
\begin{center}
\begin{tabular}{lllllllll}
& & & & $\ba{c}(10,10)\\\B_{\dalpha\dbeta\alpha\beta}\ea\!$ & & & &  \cr
& & & $\diagup$ & & $\,\diagdown$ & & & \cr
& & $\ba{c}(36,4)\\\B^{\alpha}{}_{\dalpha\beta\gamma}\ea\!$ & & & & $\ba{c}(4,36)\\\B_{\dalpha\dbeta}{}^{\dgamma}{}_{\alpha}\ea\!$& &  \cr 
& $\diagup$ & & $\diagdown$ & & $\diagup$ & & $\diagdown$ & \cr
$\ba{c}(45,1)\\\B^{\alpha\beta}{}_{\gamma\delta}\ea\!$ & & & & $\ba{c}(15,15)\\\B^{\alpha}{}_{\dbeta}{}^{\dalpha}{}_{\beta}\ea\!$& & & &
$\ba{c}(1, 45)\\\B_{\dalpha\dbeta}{}^{\dgamma\ddelta}\ea\!$\cr
& $\diagdown$ & & $\diagup$ & & $\diagdown$ & & $\diagup$ & \cr
& & $\ba{c}(20,\bar{4})\\\B^{\alpha\beta\dalpha}{}_{\gamma}\ea\!$ & & & & $\ba{c}(\bar{4},20)\\\B^{\alpha}{}_{\dalpha}{}^{\dbeta\dgamma}\ea\!$ & &  \cr  
& & & $\diagdown$ & & $\diagup$ & & & \cr
& & & & $~\ba{c}(6,6)\\\B^{\alpha\beta\dalpha\dbeta}\ea\!$ & & & & 
\end{tabular}
\end{center}
where the diagonal lines link the two neighboring $\su(2,2)\oplus\su(4)$ multiplets which are connected  by the supercharges. The complex dimension of the supermultiplet is $899$. 
 Below we explicitly  present all the super-commutation relations of the extended superalgebra,
\be
\ba{l}
{}[Q_{\kappa\dkappa},\B^{\alpha\beta\dalpha\dbeta}]=
-\textstyle{\frac{1}{3}}\delta_{\kappa}^{\,\alpha}\B^{\beta}{}_{\dkappa}{}^{\dalpha\dbeta}
+\textstyle{\frac{1}{3}}\delta_{\kappa}^{\,\beta}\B^{\alpha}{}_{\dkappa}{}^{\dalpha\dbeta}
-\textstyle{\frac{1}{3}}\delta_{\dkappa}^{\,\dalpha}\B^{\alpha\beta\dbeta}{}_{\kappa}
+\textstyle{\frac{1}{3}}\delta_{\dkappa}^{\,\dbeta}\B^{\alpha\beta\dalpha}{}_{\kappa}\,,\\
{}\\
{}\{Q_{\kappa\dkappa},\B^{\alpha}{}_{\dalpha}{}^{\dbeta\dgamma}\}=
\textstyle{\frac{3}{8}}\delta_{\dkappa}^{\,\dbeta}\B^{\alpha}{}_{\dalpha}{}^{\dgamma}{}_{\kappa}
-\textstyle{\frac{3}{8}}\delta_{\dkappa}^{\,\dgamma}\B^{\alpha}{}_{\dalpha}{}^{\dbeta}{}_{\kappa}
-\textstyle{\frac{1}{8}}\delta_{\dalpha}^{\,\dbeta}\B^{\alpha}{}_{\dkappa}{}^{\dgamma}{}_{\kappa}
+\textstyle{\frac{1}{8}}\delta_{\dalpha}^{\,\dgamma}\B^{\alpha}{}_{\dkappa}{}^{\dbeta}{}_{\kappa}
+\textstyle{\frac{1}{4}}\delta_{\kappa}^{\,\alpha}\B_{\dkappa\dalpha}{}^{\dbeta\dgamma}\,,\\
{}\\
{}\{Q_{\kappa\dkappa},\B^{\alpha\beta\dalpha}{}_{\gamma}\}=
-\textstyle{\frac{3}{8}}\delta_{\kappa}^{\,\alpha}\B^{\beta}{}_{\dkappa}{}^{\dalpha}{}_{\gamma}
+\textstyle{\frac{3}{8}}\delta_{\kappa}^{\,\beta}\B^{\alpha}{}_{\dkappa}{}^{\dalpha}{}_{\gamma}
+\textstyle{\frac{1}{8}}\delta_{\gamma}^{\,\alpha}\B^{\beta}{}_{\dkappa}{}^{\dalpha}{}_{\kappa}
-\textstyle{\frac{1}{8}}\delta_{\gamma}^{\,\beta}\B^{\alpha}{}_{\dkappa}{}^{\dalpha}{}_{\kappa}
+\textstyle{\frac{1}{4}}\delta_{\dkappa}^{\,\dalpha}\B^{\alpha\beta}{}_{\gamma\kappa}\,,
\\
{}\\
{}[Q_{\kappa\dkappa},\B^{\alpha}{}_{\dbeta}{}^{\dalpha}{}_{\beta}]=
\textstyle{\frac{4}{15}}\delta_{\kappa}^{\,\alpha}\B_{\dkappa\dbeta}{}^{\dalpha}{}_{\beta}
-\textstyle{\frac{1}{15}}\delta_{\beta}^{\,\alpha}\B_{\dkappa\dbeta}{}^{\dalpha}{}_{\kappa}
-\textstyle{\frac{4}{15}}\delta_{\dkappa}^{\,\dalpha}\B^{\alpha}{}_{\dbeta\kappa\beta}
+\textstyle{\frac{1}{15}}\delta_{\dbeta}^{\,\dalpha}\B^{\alpha}{}_{\dkappa\kappa\beta}\,,
\\
{}\\
{}[Q_{\kappa\dkappa},\B_{\dalpha\dbeta}{}^{\dgamma\ddelta}]\!=\!
\textstyle{\frac{2}{5}}\delta_{\dkappa}^{\,\ddelta}\B_{\dalpha\dbeta}{}^{\dgamma}{}_{\kappa}
\!-\!\textstyle{\frac{2}{5}}\delta_{\dkappa}^{\,\dgamma}\B_{\dalpha\dbeta}{}^{\ddelta}{}_{\kappa}
\!+\!\textstyle{\frac{1}{10}}\delta_{\dalpha}^{\,\dgamma}\B_{\dkappa\dbeta}{}^{\ddelta}{}_{\kappa}
\!+\!\textstyle{\frac{1}{10}}\delta_{\dbeta}^{\,\dgamma}\B_{\dkappa\dalpha}{}^{\ddelta}{}_{\kappa}
\!-\!\textstyle{\frac{1}{10}}\delta_{\dalpha}^{\,\ddelta}\B_{\dkappa\dbeta}{}^{\dgamma}{}_{\kappa}
\!-\!\textstyle{\frac{1}{10}}\delta_{\dbeta}^{\,\ddelta}
\B_{\dkappa\dalpha}{}^{\dgamma}{}_{\kappa},\\
{}\\
{}[Q_{\kappa\dkappa},\B^{\alpha\beta}{}_{\gamma\delta}]=
-\textstyle{\frac{2}{5}}\delta_{\kappa}^{\,\alpha}\B^{\beta}_{\dkappa\gamma\delta}
+\textstyle{\frac{2}{5}}\delta_{\kappa}^{\,\beta}\B^{\alpha}_{\dkappa\gamma\delta}
+\textstyle{\frac{1}{10}}\delta_{\gamma}^{\,\alpha}\B^{\beta}_{\dkappa\delta\kappa}
+\textstyle{\frac{1}{10}}\delta_{\delta}^{\,\alpha}\B^{\beta}_{\dkappa\gamma\kappa}
-\textstyle{\frac{1}{10}}\delta_{\gamma}^{\,\beta}\B^{\alpha}_{\dkappa\delta\kappa}
-\textstyle{\frac{1}{10}}\delta_{\delta}^{\,\beta}\B^{\alpha}_{\dkappa\gamma\kappa}\,,\\
{}\\
{}\{Q_{\kappa\dkappa},\B^{\alpha}{}_{\dalpha\beta\gamma}\}=
\textstyle{\frac{5}{18}}\delta_{\kappa}^{\,\alpha}\B_{\dkappa\dalpha\beta\gamma}
-\textstyle{\frac{1}{18}}\delta_{\beta}^{\,\alpha}\B_{\dkappa\dalpha\kappa\gamma}
-\textstyle{\frac{1}{18}}\delta_{\gamma}^{\,\alpha}\B_{\dkappa\dalpha\kappa\beta}\,,\\
{}\\
{}\{Q_{\kappa\dkappa},\B_{\dalpha\dbeta}{}^{\dgamma}{}_{\alpha}\}=
\textstyle{\frac{5}{18}}\delta_{\dkappa}^{\,\dgamma}\B_{\dalpha\dbeta\kappa\alpha}
-\textstyle{\frac{1}{18}}\delta_{\dalpha}^{\,\dgamma}\B_{\dkappa\dbeta\kappa\alpha}
-\textstyle{\frac{1}{18}}\delta_{\dbeta}^{\,\dgamma}\B_{\dkappa\dalpha\kappa\alpha}\,,\\
{}\\
{}[Q_{\kappa\dkappa},\B_{\dalpha\dbeta\alpha\beta}]=0\,.
\ea\label{MAIN1}
\ee

All the brane charges are traceless, anti-symmetric for the upper indices, and symmetric for the lower indices if they belong to the same species.  The statistics of the brane charges depends  whether the number of the upper indices is even or odd. Furthermore, the upper  index can be lowered and converted to  the  different species using the positive supercharges, $Q_{\alpha\dalpha}$, from right to left. For example, 
\be
\ba{ll}
{}[Q_{\gamma\dvarepsilon}\,,\,B^{\alpha\beta\dalpha\dvarepsilon}]
=B^{\alpha\beta\dalpha}{}_{\gamma}=B^{[\alpha\beta]\dalpha}{}_{\gamma}\,,~~~&~~~
[Q_{\varepsilon\dgamma}\,,\,B^{\alpha\varepsilon\dalpha\dbeta}]
=B^{\alpha}{}_{\dgamma}{}^{\dalpha\dbeta}=B^{\alpha}{}_{\dgamma}{}^{[\dalpha\dbeta]}\,,\\
{}&{}\\
{}\{Q_{\varepsilon\dbeta}\,,\,B^{\alpha\varepsilon\dalpha}{}_{\beta}\}=
B^{\alpha}{}_{\dbeta}{}^{\dalpha}{}_{\beta}\,,~~~&~~~
{}\{Q_{\varepsilon\dalpha}\,,\,B^{\varepsilon}{}_{\dbeta}{}^{\dgamma\ddelta}\}=
B_{\dalpha\dbeta}{}^{\dgamma\ddelta}=B_{(\dalpha\dbeta)}{}^{[\dgamma\ddelta]}\,,\\
{}&{}\\
{}\{Q_{\gamma\dvarepsilon}\,,\,B^{\alpha\beta\dvarepsilon}{}_{\delta}\}
=B^{\alpha\beta}{}_{\gamma\delta}=B^{[\alpha\beta]}{}_{(\gamma\delta)}\,,~~~&~~~
[Q_{\varepsilon\dalpha}\,,\,B^{\alpha\varepsilon}{}_{\beta\gamma}]
=B^{\alpha}{}_{\dalpha\beta\gamma}=B^{\alpha}{}_{\dalpha(\beta\gamma)}\,,\\
{}&{}\\
{}[Q_{\varepsilon\dalpha}\,,\,B^{\varepsilon}{}_{\dbeta}{}^{\dgamma}{}_{\alpha}]=
B_{\dalpha\dbeta}{}^{\dgamma}{}_{\alpha}=
B_{(\dalpha\dbeta)}{}^{\dgamma}{}_{\alpha}\,,~~~&~~~
{}\{Q_{\varepsilon\dalpha}\,,\,B^{\varepsilon}{}_{\dbeta\alpha\beta}\}=
B_{\dalpha\dbeta\alpha\beta}=
B_{(\dalpha\dbeta)(\alpha\beta)}\,.
\ea
\ee
Note that the tracelessness  follows from (\ref{Nconstraint}).

The super-commutators  between the negative supercharges and the brane charges can be also obtained recurrently, floor by floor, using the above expressions  for the brane charges and the superalgebra itself, (\ref{D1D3sa}). They are 
\be
\ba{l}
{}[\bar{Q}^{\kappa\dkappa},\B_{\dalpha\dbeta\alpha\beta}]=
\textstyle{\frac{72}{5}}\delta_{\,\alpha}^{\kappa}\B_{\dalpha\dbeta}{}^{\dkappa}{}_{\beta}
+\textstyle{\frac{72}{5}}\delta_{\,\beta}^{\kappa}\B_{\dalpha\dbeta}{}^{\dkappa}{}_{\alpha}
-\textstyle{\frac{72}{5}}\delta_{\,\dalpha}^{\dkappa}\B^{\kappa}{}_{\dbeta\alpha\beta}
-\textstyle{\frac{72}{5}}\delta_{\,\dbeta}^{\dkappa}\B^{\kappa}{}_{\dalpha\alpha\beta}\,,\\
{}\\
{}\{\bar{Q}^{\kappa\dkappa},\B^{\alpha}{}_{\dalpha\beta\gamma}\}=
10\delta^{\dkappa}_{\,\dalpha}\B^{\kappa\alpha}{}_{\beta\gamma}
-15\delta^{\kappa}_{\,\beta}\B^{\alpha}{}_{\dalpha}{}^{\dkappa}{}_{\gamma}
-15\delta^{\kappa}_{\,\gamma}\B^{\alpha}{}_{\dalpha}{}^{\dkappa}{}_{\beta}
+3\delta^{\alpha}_{\,\beta}\B^{\kappa}{}_{\dalpha}{}^{\dkappa}{}_{\gamma}
+3\delta^{\alpha}_{\,\gamma}\B^{\kappa}{}_{\dalpha}{}^{\dkappa}{}_{\beta}\,,\\
{}\\
{}\{\bar{Q}^{\kappa\dkappa},\B_{\dalpha\dbeta}{}^{\dgamma}{}_{\alpha}\}=
10\delta^{\kappa}_{\,\alpha}\B_{\dalpha\dbeta}{}^{\dgamma\dkappa}
-15\delta^{\dkappa}_{\,\dalpha}\B^{\kappa}{}_{\dbeta}{}^{\dgamma}{}_{\alpha}
-15\delta^{\dkappa}_{\,\dbeta}\B^{\kappa}{}_{\dalpha}{}^{\dgamma}{}_{\alpha}
+3\delta_{\,\dalpha}^{\dgamma}\B^{\kappa}{}_{\dbeta}{}^{\dkappa}{}_{\alpha}
+3\delta_{\,\dbeta}^{\dgamma}\B^{\kappa}{}_{\dalpha}{}^{\dkappa}{}_{\alpha}\,,\\
{}\\
{}[\bar{Q}^{\kappa\dkappa},\B^{\alpha}{}_{\dbeta}{}^{\dalpha}{}_{\beta}]=
-\textstyle{\frac{32}{3}}\delta^{\dkappa}_{\,\dbeta}\B^{\alpha\kappa\dalpha}{}_{\beta}
-\textstyle{\frac{32}{3}}\delta^{\kappa}_{\,\beta}\B^{\alpha}{}_{\dbeta}{}^{\dalpha\dkappa}
+\textstyle{\frac{8}{3}}\delta^{\dalpha}_{\,\dbeta}\B^{\alpha\kappa\dkappa}{}_{\beta}
+\textstyle{\frac{8}{3}}\delta^{\alpha}_{\,\beta}\B^{\kappa}{}_{\dbeta}{}^{\dalpha\dkappa}\,,\\
{}\\
{}[\bar{Q}^{\kappa\dkappa},\B_{\dalpha\dbeta}{}^{\dgamma\ddelta}]\!=\!
-16\delta^{\dkappa}_{\,\dalpha}\B^{\kappa}{}_{\dbeta}{}^{\dgamma\ddelta}
-\!16\delta^{\dkappa}_{\,\dbeta}\B^{\kappa}{}_{\dalpha}{}^{\dgamma\ddelta}
-\!4\delta^{\dgamma}_{\,\dalpha}\B^{\kappa}{}_{\dbeta}{}^{\ddelta\dkappa}
-\!4\delta^{\dgamma}_{\,\dbeta}\B^{\kappa}{}_{\dalpha}{}^{\ddelta\dkappa}
\!+4\delta^{\ddelta}_{\,\dalpha}\B^{\kappa}{}_{\dbeta}{}^{\dgamma\dkappa}
\!+4\delta^{\ddelta}_{\,\dbeta}\B^{\kappa}{}_{\dalpha}{}^{\dgamma\dkappa}\,,\\
{}\\
{}[\bar{Q}^{\kappa\dkappa},\B^{\alpha\beta}{}_{\gamma\delta}]=
16\delta^{\kappa}_{\,\gamma}\B^{\alpha\beta\dkappa}{}_{\delta}
+16\delta^{\kappa}_{\,\delta}\B^{\alpha\beta\dkappa}{}_{\gamma}
-4\delta^{\alpha}_{\,\gamma}\B^{\kappa\beta\dkappa}{}_{\delta}
-4\delta^{\alpha}_{\,\delta}\B^{\kappa\beta\dkappa}{}_{\gamma}
+4\delta^{\beta}_{\,\gamma}\B^{\kappa\alpha\dkappa}{}_{\delta}
+4\delta^{\beta}_{\,\delta}\B^{\kappa\alpha\dkappa}{}_{\gamma}\,,\\
{}\\
{}\{\bar{Q}^{\kappa\dkappa},\B^{\alpha}{}_{\dgamma}{}^{\dalpha\dbeta}\}=
-12\delta^{\dkappa}_{\,\dgamma}\B^{\alpha\kappa\dalpha\dbeta}
-4\delta^{\dalpha}_{\,\dgamma}\B^{\alpha\kappa\dbeta\dkappa}
+4\delta^{\dbeta}_{\,\dgamma}\B^{\alpha\kappa\dalpha\dkappa}\,,\\
{}\\
{}\{\bar{Q}^{\kappa\dkappa},\B^{\alpha\beta\dalpha}{}_{\gamma}\}=
12\delta^{\kappa}_{\,\gamma}\B^{\alpha\beta\dalpha\dkappa}
+4\delta^{\alpha}_{\,\gamma}\B^{\beta\kappa\dalpha\dkappa}
-4\delta^{\beta}_{\,\gamma}\B^{\alpha\kappa\dalpha\dkappa}\,,\\
{}\\
{}[\bar{Q}^{\kappa\dkappa},\B^{\alpha\beta\dalpha\dbeta}]=0\,.
\ea\label{MAIN2}
\ee

Note that the $D1$ brane charge, $Z_{\mu a}$, as well as the top floor brane charge, $\B_{\dalpha\dbeta\alpha\beta}$, are annihilated by eight real supercharges, which shows that the adjoint supermultiplet formed by the brane charges is ``$8/32$ BPS multiplet". \newline
{}\newline
{}\newline


\newpage

\section{Comments\label{comments}}
\subsection{Translation to the  $\CN=4$ superalgebra in four dimensions\label{fourD}}
In terms of the twelve dimensional conventions  introduced in Section \ref{setting}, the fully extended $AdS_{5}\times S^{5}$ superalgebra, (\ref{QQG}), reads 
\be
\ba{ll}
\{\CQ,\bar{\CQ}\}=P_{{\13}}\!\Big[&
i\Gamma^{\mu\nu}M_{\mu\nu}
-i\Gamma^{ab}M_{ab}
+\textstyle{\frac{1}{4}}\Gamma^{\mu\nu ab}\Gamma^{(7)}\sR_{\mu\nu ab}
+P_{\7}\Gamma^{\mu a}Z_{\mu a}-\Gamma^{\mu a}P_{\7}Z_{\mu a}{}^{\dagger}\\
{}&{}\\
{}&~+\textstyle{\frac{1}{144}}P_{\7}\Gamma^{\mu\nu\lambda abc}Z^{-}_{\mu\nu\lambda abc}
-\textstyle{\frac{1}{144}}\Gamma^{\mu\nu\lambda abc}P_{\7}Z^{-}_{\mu\nu\lambda abc}{}^{\dagger}\Big]\!P_{\13}\,,
\ea\label{MAIN3}
\ee
where $\Gamma^{(7)}=i\Gamma^{123456}$ and $P_{\7}=\half(1+\Gamma^{(7)})$.\newline

In order to translate our results to  the four dimensional language, we need to write all  the higher dimensional objects in terms of the four dimensional conventions.  For the gamma matrices we refer (\ref{4Dsc}) in  Appendix. For the $\mbox{so}(2,4)$ generators we decompose them into the four dimensional Lorentz generators, $\hat{M}_{mn}$,  momenta, $P_{m}$, special conformal transformation generators, $K_{m}$ and Dilation, $D$, with $m,n=0,1,2,3$,
\be
\ba{llll}
\hat{M}_{mn}=M_{2+m\,2+n}\,,&~P_{m}=-M_{1\,m+2}+M_{m+2\,6}\,,&~
K_{m}=M_{1\,m+2}+M_{m+2\,6}\,,&~D=M_{1\,6}\,.
\ea
\ee
The twelve dimensional Majorana-Weyl  supercharge, $\CQ$, consists of  the four dimensional ordinary supercharges, $q$, $\bar{q}=q^{\dagger}$, and the conformal supercharges, $s$, $\bar{s}=s^{\dagger}$. As they have the opposite mass dimensions,  each of them can be singled out by the projection operator, $\half(1\mp\Gamma_{16})$. In our choice of the gamma matrices (\ref{4Dsc}), $Q_{1\dalpha}, Q_{2\dalpha}, \bar{Q}^{3\dalpha}, \bar{Q}^{4\dalpha}$ correspond to the ordinary supercharges so that
\be
\ba{cc}
Q_{\alpha\dalpha}=(\,q_{1\dalpha}\,,\,q_{2\dalpha}\,,\,
-i\bar{s}^{1}{}_{\dalpha}\,,\,-i\bar{s}^{2}{}_{\dalpha}\,)^{t}\,,~~~&~~~
\bar{Q}^{\alpha\dalpha}=(\,s^{1\dalpha}\,,\,s^{2\dalpha}\,,\,
i\bar{q}_{1}{}^{\dalpha}\,,\,i\bar{q}_{2}{}^{\dalpha}\,)\,.
\ea
\ee

Provided the above dictionary, our extended $AdS_{5}\times S^{5}$ superalgebra, (\ref{D1D3sa}),  (\ref{MAIN1}),  (\ref{MAIN2}), leads to a  noncentral extension of the four dimensional  $\CN=4$ superconformal algebra.\footnote{Our  conventions have been chosen to agree with  \cite{JHP4D} for the unextended sector.}

\subsection{On super Yang-Mills theory and more}
In the standard approach to the $\CN=4$ super Yang-Mills theory, different  vacuum expectation values (vev) of the Higgs correspond to the different theory. Especially  for the nonzero values, the conformal symmetry is spontaneously broken, and the Hilbert space parameterized by the Higgs vevs is not invariant under the conformal generators.\footnote{Strictly speaking, this is for the super-Yang-Mills theory on $R^{3,1}$.  For the theories on compact spaces, one should integrate over different  vevs of  the Higgs due to the normalizability of the zero modes.}   The truncation of our extended $\su(2,2|4)$ superalgebra to an extended  four dimensional   $\CN=4$ super Poincar\'{e} algebra can be achieved by the projection operator, $\half(1-\Gamma_{16})$. Essentially the  extended super Poincar\'{e} algebra reads, in terms of the ten dimensional gamma matrices, (\ref{App10}), and Majorana-Weyl supercharge,  (\ref{App10Q}), 
\be
\ba{ll}
{}\{\hat{Q},\bar{\hat{Q}}\}=2P_{+}\Big[&\hat{\Gamma}^{m}P_{m}+
\half\hat{\Gamma}_{(5)}\hat{\Gamma}^{mab}H_{mab}+\hat{\Gamma}^{a}T^{e}_{a}+
\hat{\Gamma}_{(5)}\hat{\Gamma}^{a}T^{g}_{a}\\
{}&{}\\
{}&~
+\textstyle{\frac{1}{24}}\hat{\Gamma}^{mnabc}\,T^{e-}_{mnabc}
+\textstyle{\frac{1}{24}}\hat{\Gamma}_{(5)}\hat{\Gamma}^{mnabc}\,T^{g-}_{mnabc}\Big]P_{-}\,,
\ea
\ee
where $\bar{\hat{Q}}=\hat{Q}^{\dagger}\hat{\Gamma}_{0}$, $P_{\pm}=\half(1\pm\hat{\Gamma}^{(11)})$, $\hat{\Gamma}_{(5)}=\hat{\Gamma}_{0123}$, and all the brane charges are real having the origin,
\be
\ba{ll}
Z_{1a}+Z_{6a}=2(T^{g}_{a}-iT^{e}_{a})\,,~~&~~
Z^{-}_{1\,m+2\,n+2\,abc}=2(T^{g-}_{mnabc}-iT^{e-}_{mnabc})\,,\\
{}&{}\\
\multicolumn{2}{c}{H_{mab}=\half(H_{1\,m+2\,ab}+H_{6\,m+2\,ab})\,.}
\ea
\ee

In particular, Osborn identified $T^{g}_{a}$ and $T^{e}_{a}$ as the electric\footnote{This  electric charge should not be confused as the gauge symmetry Noether charge. The latter  is given by the Gauss' law or the equation of motion for $A_{0}$.} and magnetic charges by investigating the supersymmetry transformation of the super-current in $\CN=4$ super Yang-Mills theory~\cite{Osborn:tq}
\be
\ba{ll}
T^{g}_{a}=\int{\rm d}{\vec{S}\cdot}\tr(\vec{B}\Phi_{a})\,,~~~&~~~~T^{e}_{a}=\int{\rm d}{\vec{S}\cdot}\tr(\vec{E}\Phi_{a})\,.
\ea
\ee
Straightforward    manipulation  can show that the \textit{ordinary} supersymmetry transformation of the electro-magnetic charges do not vanish even at the on-shell level.\footnote{Even Eq.(\ref{zeroth}) does not hold in general. This seems to imply that  the  expression of  $Z_{\mu a}$ further decomposes into several sectors which belong to different irreducible representations corresponding to various  configurations,  $(D1,D3)$, $(D1,D3,D5)$,  $(D1,D3,D3)$, etc.} Our results, (\ref{MAIN1}) and (\ref{MAIN2}), also confirm this, since the brane charges on the ground floor of the adjoint supermultiplet are annihilated by eight real supercharges out of 32.  Surprisingly this  means the noncentral property of the electro-magnetic charge, in contrast to the conventional wisdom due to the Haag-Lopuszanski-Sohnius theorem~\cite{Haag:1974qh}. The original argument for the electro-magnetic charge to be central is based on the  Coleman-Mandula theorem~\cite{Coleman:ad} on all the possible   \textit{symmetry generators} in the quantum field theories. The point for the brane charges we discuss in the paper is that they are not symmetry generators nor Noether charges. Rather, they are  topological living  at the spatial infinity only, and hence free from the constraint by the Haag-Lopuszanski-Sohnius theorem. \newline

Nevertheless, for the ordinary supersymmetric monopole configurations, our new brane charges,  at least for those coming from the ordinary supercharges, annihilate the corresponding   quantum states as follows.  Although the classical monopole or solitons are given by the bosonic configurations only, at the quantum level the fermions act nontrivially on the quantum states essentially  to respect the  second quantization of them. In other words, there is no quantum state which is annihilated by all the fermions, and one should always keep in mind the fermions. Now for the supersymmetric monopoles, the fermionic zero modes are given by the broken ordinary supersymmetry transformations of the gauginos, $\lambda\sim F_{AB}\hat{\Gamma}^{AB}\varepsilon$. The expressions for the new brane charges coming from the ordinary supercharges contain  the gauginos, the field strengths, and the derivatives of the Higgs,  but not the Higgs itself, so that, from the asymptotic behavior, one can expect that   the corresponding  new brane charges annihilate the monopole states. \newline

 It will be very interesting to find out  novel configurations which have nontrivial realization of the new brane charges, either on the super Yang-Mills side or on the supergravity side. In the former case, the full expressions for the brane charges coming from all the  ordinary as well as the conformal supercharges are desirable, which deserves a  separate analysis. Certainly, nonvanishing vevs of any brane charge imply  the dynamical breaking of  supersymmetry~\cite{Witten:nf}.  Another thing to be done  is to classify the representations of the extended $AdS$ superalgebra as in \cite{Dobrev,Dolan:2002zh}. More detailed study of the extended  superalgebra may shed light on the nonperturbative aspects of the  string/M-theory on the $AdS_{5}\times S^{5}$ background. \newline
\\
\\
\acknowledgments{JHP would like to thank    N. Dorey, S. Krusch,  A. Losev, N. Nekrasov and H. Osborn for the helpful comments and the enlightening discussions.} 
\newline
\newline
\newpage
\appendix
\section{Decomposition of the gamma matrices for lower dimensions}
\subsection{For the four dimensional  $\CN=4$ superconformal algebra \label{Appendix4D}}
In order to translate our results to  the four dimensional language, we need to write all  the higher dimensional objects in terms of the four dimensional conventions.   
First we let  the  six dimensional gamma matrices satisfying (\ref{gamma7}) and (\ref{herm}) be
\be
\ba{lll}
\rho_{1}=\left(\ba{cc}0&+1\\+1&0\ea\right)\,,~~~&~~~
\rho_{m+2}=\left(\ba{cc}{\sigma}_{m}&0\\0&\bar{\sigma}_{m}\ea\right)\,,~~~&~~~
\rho_{6}=\left(\ba{cc}0&-1\\+1&0\ea\right)\,,\\
{}&{}&{}\\
\bar{\rho}_{1}=\left(\ba{cc}0&-1\\-1&0\ea\right)\,,~~~&~~~
\bar{\rho}_{m+2}=\left(\ba{cc}\bar{\sigma}_{m}&0\\0&{\sigma}_{m}\ea\right)\,,~~~&~~~
\bar{\rho}_{6}=\left(\ba{cc}0&+1\\-1&0\ea\right)\,,
\ea\label{4Dsc}
\ee
where the $2\times 2$ matrices, $\sigma_{m}=(+1,\vec{\tau})$, $\bar{\sigma}_{m}=(-1,\vec{\tau})$, $m=0,1,2,3$, satisfy the Clifford algebra  of the four dimensional spacetime on which the super Yang-Mills exists,
\be
\ba{cc}
\sigma_{m}\bar{\sigma}_{n}+\sigma_{n}\bar{\sigma}_{m}=2\hat{\eta}_{mn}\,,~~~&~~
\hat{\eta}=\mbox{diag}(-+++)\,.
\ea
\ee
The four dimensional gamma matrices are then
\be
\ba{ll}
\hat{\gamma}_{m}=\left(\ba{cc}0~&\sigma_{m}\\\bar{\sigma}_{m}&0\ea\right)
=\fB^{-1}(\hat{\gamma}_{m})^{\ast}\fB\,,~~&~\fB=
\left(\ba{cc}0~&\epsilon\\\epsilon^{-1}&0\ea\right)\,,
\ea
\ee
where $\epsilon$ is the usual $2\times 2$  anti-symmetric matrix satisfying $\sigma_{m}^{t}=\sigma_{m}^{\ast}=\epsilon\bar{\sigma}_{m}\epsilon$, $\epsilon_{{\scriptscriptstyle 12}}=1$.\newline

The above $\rho_{\mu}$ matrices are not anti-symmetric, and to make them so one needs to take some transformations  such as
\be
\ba{ll}
\rho_{\mu}~\rightarrow~\left(\ba{cc}0~&\epsilon\\\epsilon^{-1}&0\ea\right)\rho_{\mu}~
\rightarrow~U\left(\ba{cc}0~&\epsilon\\\epsilon^{-1}&0\ea\right)\rho_{\mu}U^{t}\,,~~&~~
U=\textstyle{\frac{1}{\sqrt{2}}}\left(\ba{cc}\tau_{1}&-i\tau_{1}\\\tau_{1}&~i\tau_{1}
\ea\right)\,.
\ea
\ee
The first transformation makes $\rho_{\mu}$'s  anti-symmetric, while  the next similarity transformation involving the unitary matrix, $U$,  further ensures that the  representation of the Cartan subalgebra is   diagonal,  exactly as  (\ref{cartan}). 
\newpage

\subsection{For the four dimensional super Poincar\'{e} algebra\label{AppPoincare}}
For the truncation of our extended  $AdS_{5}\times S^{5}$ superalgebra to an extended super Poincar\'{e} algebra in four dimensions, we write the twelve dimensional gamma matrices in terms of the ten  dimensional ones, $\hat{\Gamma}^{A}$, $A=m,a$,
\be
\ba{ll}
\Gamma^{1}=\epsilon\otimes 1\,,~~~&~~\Gamma^{6}=\tau^{1}\otimes 1\,,\\
{}&{}\\
\Gamma^{m+2}=\tau^{3}\otimes\hat{\Gamma}^{m}\,,
~~~&~~\Gamma^{a}=\tau^{3}\otimes\hat{\Gamma}^{a}\,.
\ea\label{App10}
\ee
Further the $10D$ gamma matrices decompose into the $4D$ and $6D$ ones,
\be
\ba{ll}
\hat{\Gamma}^{m}=\hat{\gamma}^{m}\otimes\gamma^{(7)}\,,~~&~~
\hat{\Gamma}^{a}=1\otimes\gamma^{a}\,,
\ea
\ee
satisfying 
\be
\ba{ccc}
(\hat{\Gamma}^{\hat{M}})^{\ast}=-\tB\hat{\Gamma}^{\hat{M}}\tB^{-1}\,,~~&~~
\tB=\fB\otimes\sB\,,~~&~~\sB={\small\left(\ba{cc}0&1\\1&0\ea\right)}\,.
\ea
\ee

$10D$ chirality matrix reads
\be
\hat{\Gamma}^{(11)}={\small\left(\ba{cc}1&0\\0&-1\ea\right)}\otimes\gamma^{(7)}\,.
\ee

Majorana-Weyl supercharge carries the $4D$ and $6D$ chiral indices of the same chirality, 
\be
\hat{Q}=\hat{\Gamma}^{(11)}\hat{Q}=B^{-1}\hat{Q}^{\ast}=
\left(
q_{1\dalpha}\,,\,q_{2\dalpha}\,,\,
(\epsilon^{-1}\bar{q}^{t})^{1\dalpha}\,,\,(\epsilon^{-1}\bar{q}^{t})^{2\dalpha}\right)^{t}\,.
\label{App10Q}
\ee

\end{document}